\documentclass[twocolumn,floatfix,superscriptaddress,amsmath,showpacs,showkeys,aps,prb]{revtex4}

\usepackage[final]{graphicx}
\usepackage{t1enc}
\usepackage{bm}

\usepackage{amsfonts}

\begin{document}
\bibliographystyle{apsrev}

\title{Non-equilibrium structures and slow dynamics in a two dimensional spin system with competitive long range and short range interactions}

\author{Omar Osenda}
\email{osenda@famaf.unc.edu.ar}
\affiliation{Instituto de Física de la Facultad de Matemática, Astronomía y Física (IFFAMAF-CONICET),
Universidad Nacional de Córdoba \\
Ciudad Universitaria, 5000 Córdoba, Argentina}
\author{Francisco A. Tamarit}
\email{tamarit@famaf.unc.edu.ar}\affiliation{Instituto de Física de la Facultad de Matemática, Astronomía y Física (IFFAMAF-CONICET),
Universidad Nacional de Córdoba \\
Ciudad Universitaria, 5000 Córdoba, Argentina}
\author{Sergio A. Cannas}
\email{cannas@famaf.unc.edu.ar} \affiliation{Instituto de Física de la Facultad de Matemática, Astronomía y Física (IFFAMAF-CONICET),
Universidad Nacional de Córdoba \\
Ciudad Universitaria, 5000 Córdoba, Argentina}
\date{\today}

\begin{abstract}
We introduce a lattice spin model that mimics a system of interacting particles through a short range repulsive potential and a long range attractive power law decaying potential. We perform a detailed analysis of the general equilibrium phase diagram of the model at finite temperature, showing that the only possible equilibrium phases are the ferromagnetic and the antiferromagnetic ones. We then study the non equilibrium behavior of the model after a quench to subcritical temperatures, in the antiferromagnetic region of the phase diagram  region, where the pair interaction potential behaves in the same qualitative way as in a Lennard-Jones gas. We find that, even in the absence of quenched disorder or geometric frustration,  the competition between interactions gives rise to non--equilibrium disordered structures at low enough temperatures that  strongly slow down the relaxation of the system. This non--equilibrium state presents several features characteristic of glassy systems, such as subaging, non trivial Fuctuation Dissipation relations and possible logarithmic growth of free energy barriers to coarsening.
\end{abstract}


\maketitle

\section{Introduction}

The nature of glassy magnetic states in the absence of quenched
disorder  has been object of a great deal of work\cite{BouKuMe1997},
both experimental and theoretical. Simple experimental
realizations of nondisordered systems in which glassy phases have
been found are antiferromagnets (AFM's) with {\em kagom\'e}
geometries\cite{WiHaMeMaTu1998,WiHaRiSm2000,GiStRaGaGr1997}. In particular the
presence of slow dynamics and aging effects in {\em kagom\'e}
antiferromagnets is well established\cite{WiDuViHaCa2000}. Anyway,
these are non--disordered geometrically frustrated systems, so the
understanding of the mechanisms present on the dynamics of these
systems has to deal with the effects of the involved geometry of
the kagom\'e lattice. On the other hand, glassy behavior in structural glasses appears dynamically, without any kind of imposed disorder or geometrical frustration. A prototype model for structural glasses is the Lennard-Jones binary mixture\cite{Ko1999}. One may wonder whether glassy behavior can appear in lattice spin systems sharing some of the basic features of the Lennard-Jones model, such as the competition between short range repulsive interactions (i.e., hard core) and long range attractive interactions. A simple model with those properties  is the  Ising model with competitive interactions on the
square lattice.

Consider the general lattice Hamiltonian

\begin{equation}
{\mathcal H} = J_1 \sum_{\langle i,j \rangle} \sigma_{i} \sigma_{j}
+ J_2 \sum_{(i,j)}   \frac{\sigma_{i} \sigma_{j}}{r_{ij}^3}
\label{Hamilton1}
\end{equation}
\noindent where $\sigma = \pm1$. The first sum runs over all pairs
of nearest neighbor spins on a square lattice and the second one
over all distinct pairs of spins of the lattice. $r_{ij}$ is the
distance, measured in crystal units between sites $i$ and $j$. For
$J_1<0$ and $J_2 >0$ this Hamiltonian describes an  ultrathin
magnetic film with perpendicular anisotropy in the monolayer
limit\cite{DeMaWh2000} and it has been the subject of several
theoretical studies (see Refs.\onlinecite{PiCa2007,CaMiStTa2006} and references
therein).

In this work we consider the case $J_2=-1$ (long range
ferromagnetic interactions) $J\equiv J_1/\left|J_2\right| >0$ (short range
antiferromagnetic interactions), so (\ref{Hamilton1}) reduces to the
dimensionless Hamiltonian
\begin{equation}
{\mathcal H} = J \sum_{\langle i,j \rangle} \sigma_{i} \sigma_{j}
- \sum_{(i,j)}   \frac{\sigma_{i} \sigma_{j}}{r_{ij}^3}
\label{Hamilton2}
\end{equation}

\begin{figure}
\begin{center}
\includegraphics[scale=0.38,angle=-90]{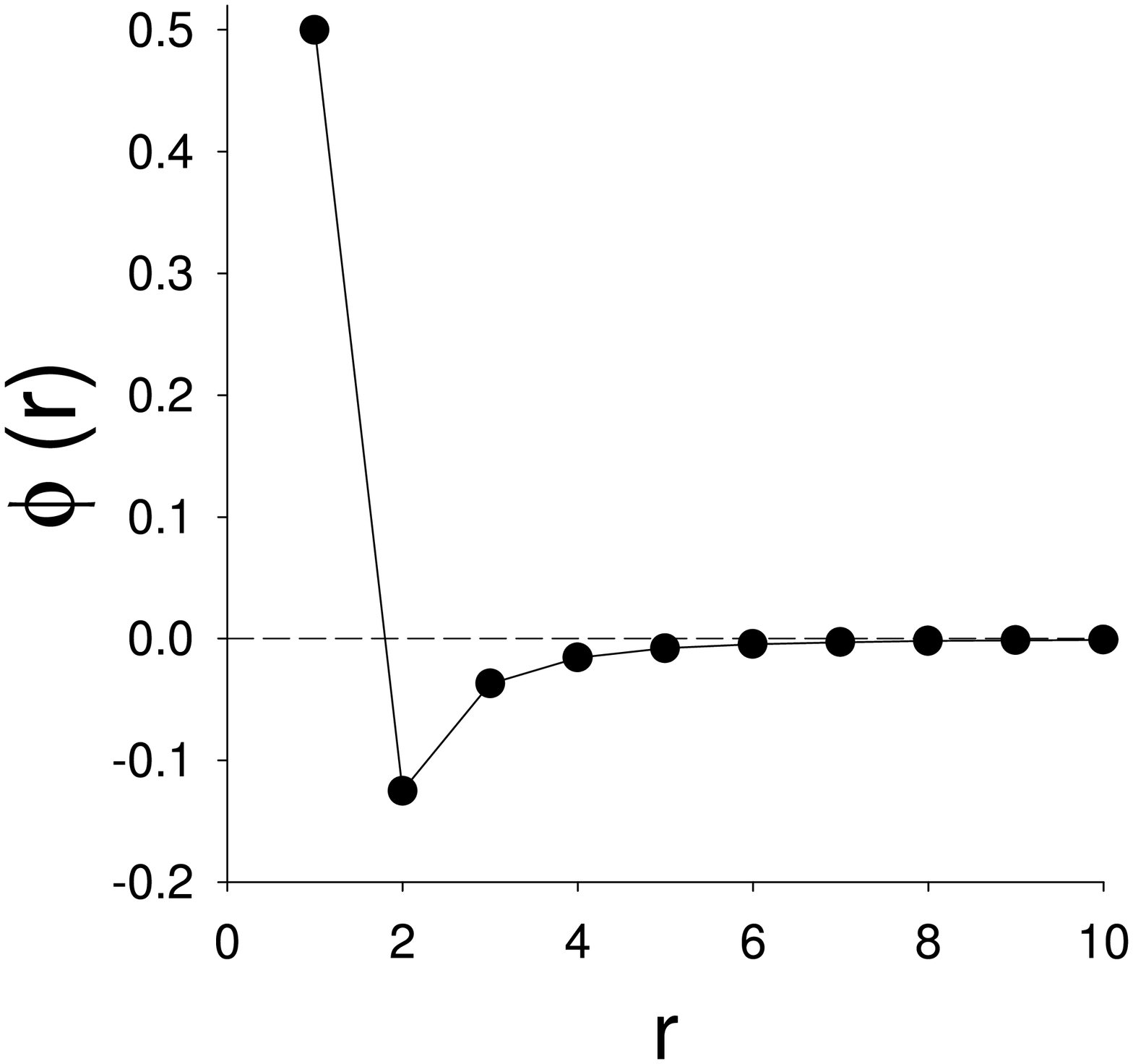}
\end{center}
\caption{Pair interaction potential as a function of distance for $J=1.5$.\label{fig1}}
\end{figure}

For $J>1$ this Hamiltonian mimics a system of particles
interacting through a short range repulsive potential and a long
range power law decaying attractive potential, qualitatively similar to the Lennard-Jones pair interactions potential (see an example in Fig.\ref{fig1}). Although the exponent $3$ in the power law
interacting potential is arbitrary in this case, it presents two
advantages. First, in a two dimensional system it is large enough
to ensure the existence of the thermodynamical limit and, second,
it allows us to use all the previous knowledge of the much more
studied related system $J<0$ and $J_1>0$ (ultrathin magnetic films
model). As we will show, it displays complex low temperature
dynamical behavior, even when its equilibrium properties are
simpler than those observed in the ultrathin magnetic films case.
In order to correlate equilibrium and non-equilibrium properties
we start our analysis by investigating the finite temperature
thermodynamical behavior of the model. Since, to the best of our
knowledge, this model has not been previously studied in the
literature and also for completeness we perform in section
\ref{equilibrium} a detailed analysis of the complete equilibrium
phase diagram using Monte Carlo (MC) simulations.  In section
\ref{nonequilibrium} we analyze the low temperature relaxation
properties of the model, by computing different quantities like
the average linear size of domains, energy, two times correlation
functions and Fluctuation-Dissipation relations. In section
\ref{conclusions} we discuss our results, comparing them with previous reported results of slow dynamics in non--disordered systems, in particular, the Lennard--Jones gas.

\section{Equilibrium phase diagram}

\label{equilibrium}

\begin{figure}
\begin{center}
\includegraphics[scale=0.4,angle=0]{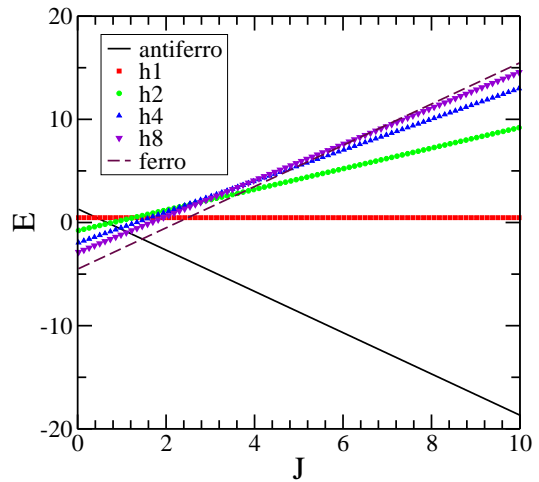}
\end{center}
\caption{(Color on-line) Energy per site {\em vs} $J$ for
different spin configuration; h1 corresponds to an state composed
by stripes with width one, h2 to stripes with width two and so on.
The chequered states give similar values for the energy per site
(not shown here) to the stripes states with the same
width.\label{figone}}
\end{figure}

 As a first step we analyze the zero temperature
properties of the model. In the absence of the long range term the
model reduces to the Ising antiferromagnetic model and the ground
state of the system is a N\'eel antiferromagnetic state. For $J=0$
it is clear that the ground state is ferromagnetic. To obtain the
ground state between these two  limits we evaluated
the energy per site for different spin configurations, namely
ferromagnetic, antiferromagnetic, stripes of different widths
($1,2,\ldots$ rows of spins) and chequered domains of different
sizes. The energies per spin of the ferromagnetic and
antiferromagnetic states are given by $E_{f}= 2J -a$ and $E_{af}=
-2J +b$ respectively, with\cite{TaGy1993} $a= 4.5168$ and
$b=1.3230$. The energy per spin of a state composed by
ferromagnetic stripes of width $h$ is given by $E_s(h) =
2\left(1-1/h \right)\, J + S_h$. The values of $S_h$ were
calculated numerically in Ref.\onlinecite{TaGy1993}; for instance
$S_1=0.4677$, $S_2=-0.7908$, etc.. In Figure \ref{figone} we
compare the energy per site for different configurations. The figure shows
that the only stable states are the ferromagnetic and the
antiferromagnetic ones for any value of $J$. By equating
$E_f=E_{af}$ we obtain for the transition point between the
ferromagnetic and the antiferromagnetic states the value
$J_t=1.4599$: the ground state is ferromagnetic below this value
and antiferromagnetic above of it. We also checked different
chequered antiferromagnetic states\cite{TaGy1993}, verifying that
they have higher energies than either the ferromagnetic or the
antiferromagnetic states for any value of $J$. Monte Carlo (MC)
simulations at low temperatures confirm that these are the only
low temperature stable phases.

We next consider the equilibrium finite temperature properties
of the model, by using Metropolis Monte Carlo algorithm in finite
square lattices  with $N=L \times L$ sites and periodic boundary
conditions. To handle the contribution of the long range terms in
the periodic boundary conditions we used the Ewald sums
technique\cite{DeMaWh2000}. In order to speed up the simulations, the codes were implemented by keeping track of all the local fields. In this way, local fields update (an operation which is of ${\cal O}(N)$) is performed only when a spin flip is accepted. This implementation is very effective when relaxation is very slow (in general at low temperatures) and therefore the acceptance rate is small, while it does not change the computational time when the acceptance rate is high (usually at intermediate or high temperatures). This will be particularly important when considering non equilibrium effects in section \ref{nonequilibrium}, allowing us to treat large system sizes at very low temperatures.

To characterize the critical
properties of the model we calculate different thermodynamical
quantities as a function of the temperature, for different values
of $J$ and $L$, namely, the magnetization per spin $m$, the
staggered magnetization per spin $m_s$, the associated
susceptibilities

\begin{equation}
\chi(T) = \frac{N}{T} (\langle m^2\rangle - \langle m\rangle^2 )
\end{equation}

\begin{equation}
\chi_s(T) = \frac{N}{T} (\langle m_s^2\rangle - \langle
m_s\rangle^2 ),
\end{equation}

\noindent the specific heat

\begin{equation}
C(T) = \frac{1}{NT^2} (\langle H^2\rangle - \langle H\rangle^2 )
\end{equation}

\noindent and the fourth order cumulant

\begin{equation}
V(T) = 1- \frac{\left< H^4 \right>}{3\left< H^2 \right>^2}.
\label{cumulant}
\end{equation}

\noindent where $\left\langle  \ldots\right\rangle $ stands for an
average over the thermal noise. All these quantities are
calculated starting from an initially equilibrated high
temperature configuration and slowly decreasing the temperature.
For every temperature the initial spin configuration is taken as
the final configuration of the previous temperature; we let the
system to equilibrate $M_1$ Monte Carlo Steps (one MCS is defined as
a complete cycle of $N$ spin update trials) and
average out the results of $M_2$ MCS, typical values of $M_1$ and $M_2$ being
around $10^5$ and $5 \times 10^5$ respectively.

\begin{figure}
\begin{center}
\includegraphics[scale=0.5,angle=0]{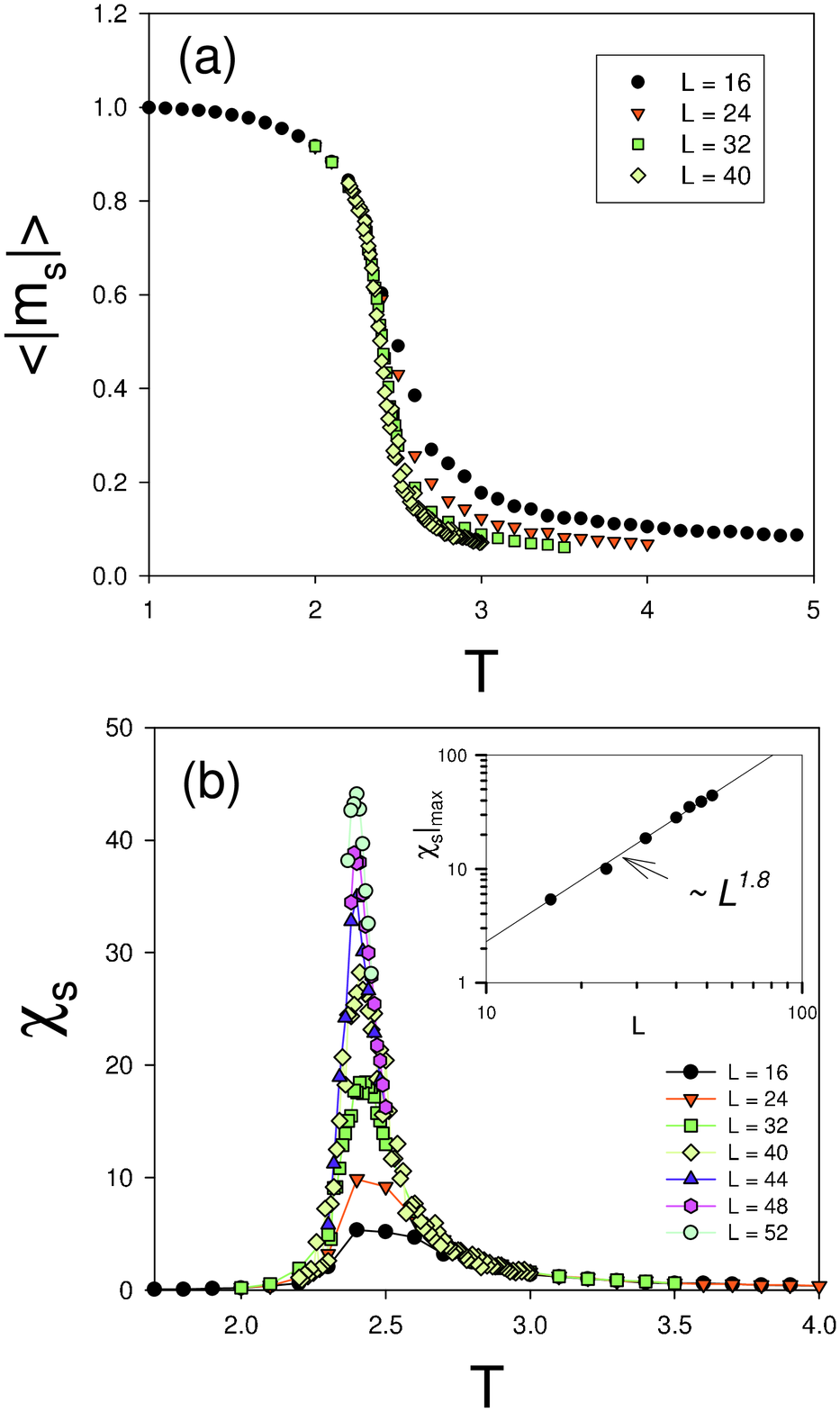}
\caption{\label{fig3} (Color on-line) Order parameter (absolute
value of the staggered magnetization) (a) and associated staggered
susceptibility (b) as a function of the temperature for $J=1.6$
and for different system sizes; the inset in (b) shows the finite
size scaling of the maximum of $\chi_s$.}
\end{center}
\end{figure}

\begin{figure}
\begin{center}
\includegraphics[scale=0.5,angle=0]{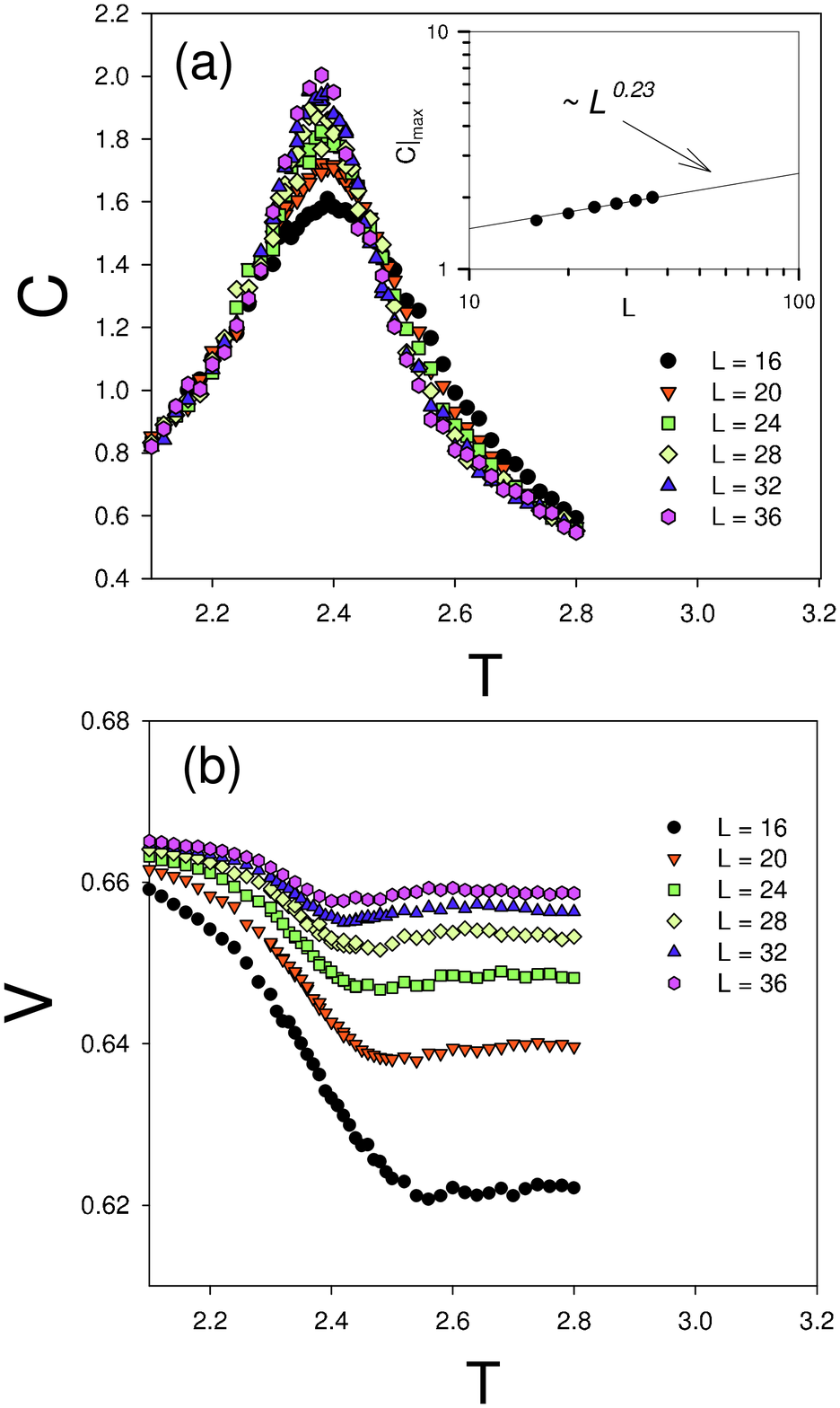}
\caption{\label{fig4} (Color on-line) Moments of the energy as a
function of the temperature for $J=1.6$ and for different system
sizes. (a) Specific heat $C$; the inset  shows the finite size
scaling of the maximum of $C$. (b) Fourth order cumulant.}
\end{center}
\end{figure}

 Figures (\ref{fig3}) and (\ref{fig4})  show the typical
results for the different thermodynamical quantities in the
antiferromagnetic region of the phase diagram, namely, for $J
>J_t$.  Fig.(\ref{fig4}b) shows that the fourth order cumulant
exhibits a vanishing minimum, consistent with a second order
phase transition. Fig.(\ref{fig3}b) shows that the staggered
susceptibility exhibits a size dependent maximum which scales as
$L^{\gamma/\nu}$, with $\gamma/\nu = 1.8 \pm 0.1$, consistent with
the exact value $\gamma/\nu=1.75$ of the two dimensional short
range Ising model. Moreover, as $J$ increases ${\gamma/\nu}$
approaches systematically the value $1.75$ (for instance, for
$J=3$ we found ${\gamma/\nu}=1.74 \pm 0.05$; see Fig.(\ref{fig7}).
We see from Fig.(\ref{fig4}a) that the specific heat exhibits a
size dependent maximum which scales as $L^{\alpha/\nu}$, with
$\alpha/\nu = 0.23 \pm 0.05$; similar values were found for other
values of $J>J_t$ (see Fig.\ref{fig7}). Although small, those
values are larger than expected for a phase transition in the
universality class of the two dimensional Ising model
($\alpha=0$). However, since those values are also observed for
large values of $J$, we believe that this is a finite size effect.
Hence, we conclude that the whole line between the paramagnetic
and the antiferromagnetic phases ($J
>J_t$) belongs to the universality class of the
short range two dimensional Ising model.

\begin{figure}
\begin{center}
\includegraphics[scale=0.5,angle=0]{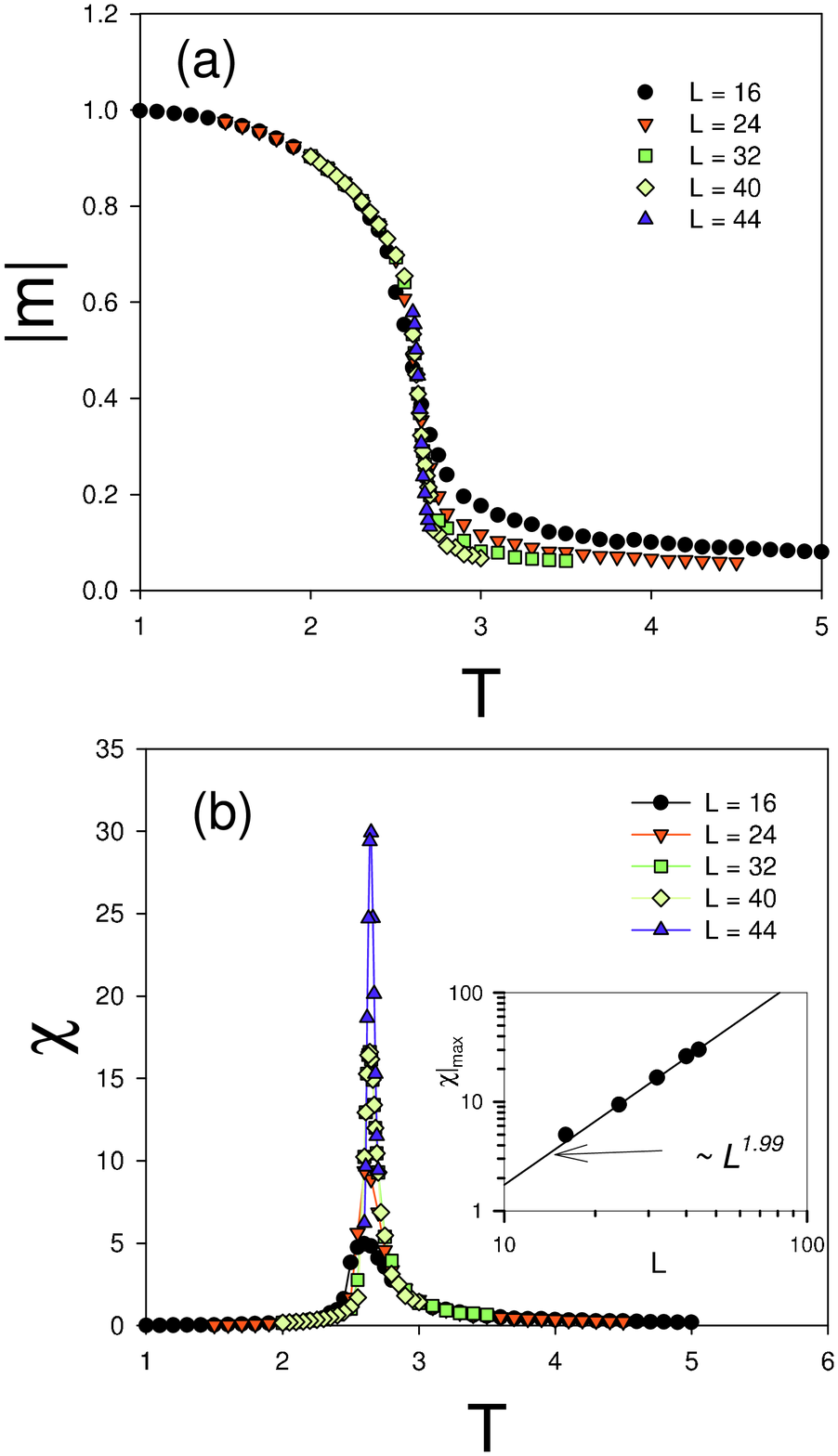}
\caption{\label{fig5} (Color on-line) Order parameter (absolute
value of the magnetization) (a) and associated susceptibility (b)
as a function of the temperature for $J=1.4$ and for different
system sizes; the inset in (b) shows the finite size scaling of
the maximum of $\chi$.}
\end{center}
\end{figure}

\begin{figure}
\begin{center}
\includegraphics[scale=0.5,angle=0]{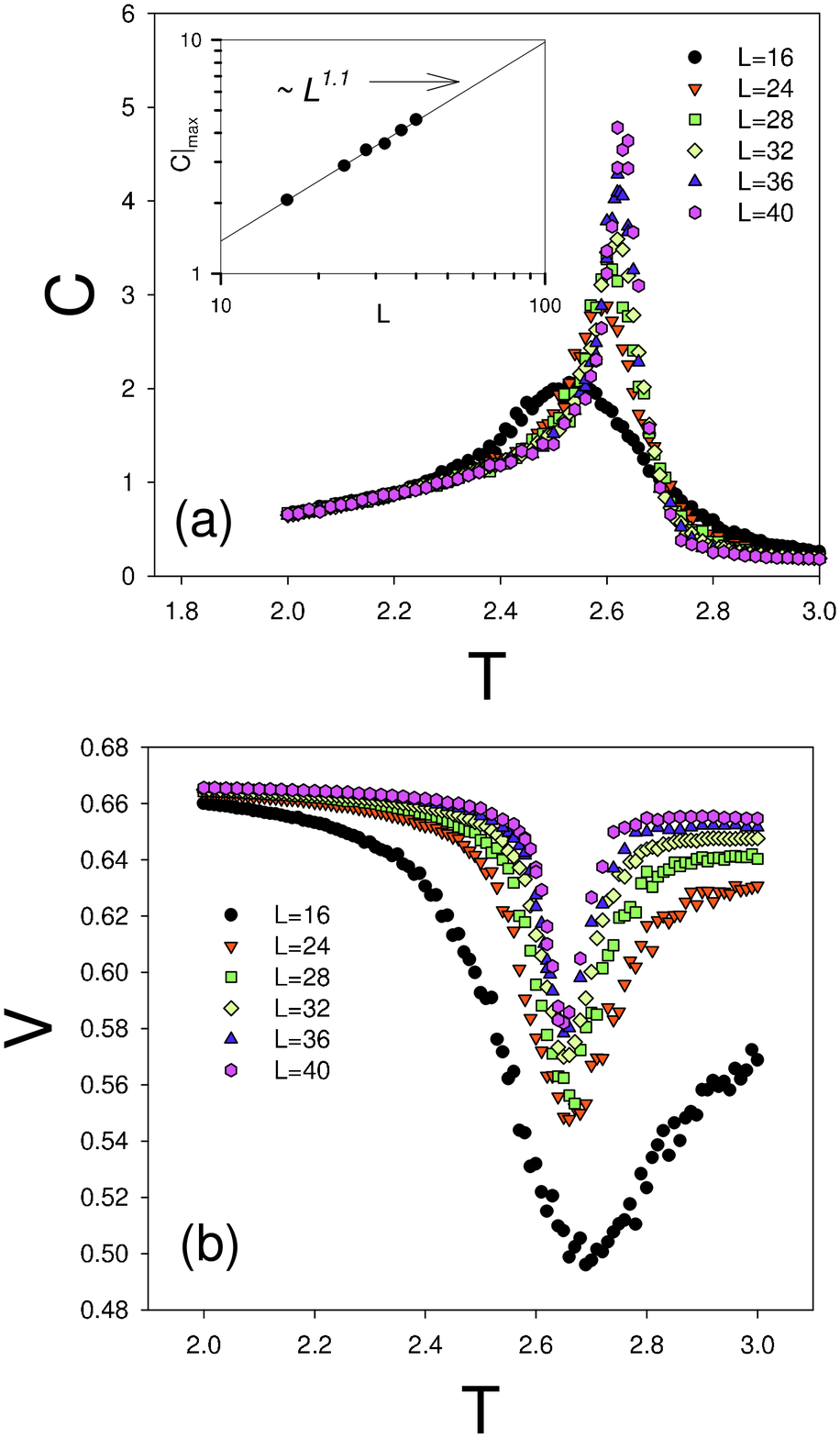}
\caption{\label{fig6} (Color on-line) Moments of the energy as a
function of the temperature for $J=1.4$ and for different system
sizes. (a) Specific heat $C$; the inset  shows the finite size
scaling of the maximum of $C$. (b) Fourth order cumulant.}
\end{center}
\end{figure}

The critical properties for $J< J_t$ are a bit more complex. For
 $J \leq 1.3$ the order parameter (magnetization), the susceptibility,
 the specific heat and the  fourth order cumulant present qualitatively
the same behavior as those quantities in the $J>J_t$ case,
but with a different set of critical exponents (see
Fig.\ref{fig7}). We found $\gamma/\nu \approx 1.1$ and $\alpha/\nu
\approx 0.14$., which are close to the renormalization group
estimates for the $J=0$ case\cite{LuBl1997}: $\nu=1$, $\alpha=0$
and $\gamma=1$; the small difference between those values and ours
can be attributed to finite size effects, which are very strong
when the long range ferromagnetic interactions dominate. Hence, we
conclude that the whole line for $0 \leq J \leq 1.3$ belongs to
the universality class of the two dimensional $1/r^3$
ferromagnetic Ising model. For $1.3< J < J_t$ we find a clear
evidence that the ferro-para transition is a first order one. The
typical behavior of the thermodynamical quantities in this case is
illustrated in Figs. \ref{fig5} and \ref{fig6}. We see that the
fourth order cumulant presents a clear converging minimum as the
system sizes increases, as expected in a first order
transition\cite{ChLaBi1986}. The finite size scaling of
susceptibility is also consistent with the $L^2$ behavior expected
for a first order transition in a two dimensional
system\cite{LeKo1991}. The specific heat exponent for $J=1.4$ is
$\alpha/\nu=1.1 \pm 0.2$. This value is certainly far from $2$
(the expected value in a first order transition), but it is larger
than the critical exponent of any continuous transition.  Besides
finite size effects, such large difference is probably also
associated to the presence of a tricritical point somewhere
between $J=1.3$ and $J=1.4$. This assumption is consistent with
the fact that $\alpha/\nu$ approaches the expected value
$\alpha/\nu=2$ as $J$ increases approaching $J=J_t$ (we obtained
$\alpha/\nu=1.8 \pm 0.1$ for $J=1.43$; see Fig.\ref{fig7}).

We summarize the obtained results for the critical exponents in
Fig.\ref{fig7} and the overall phase diagram in Fig.\ref{phased}.

\begin{figure}
\begin{center}
\includegraphics[scale=0.5,angle=0]{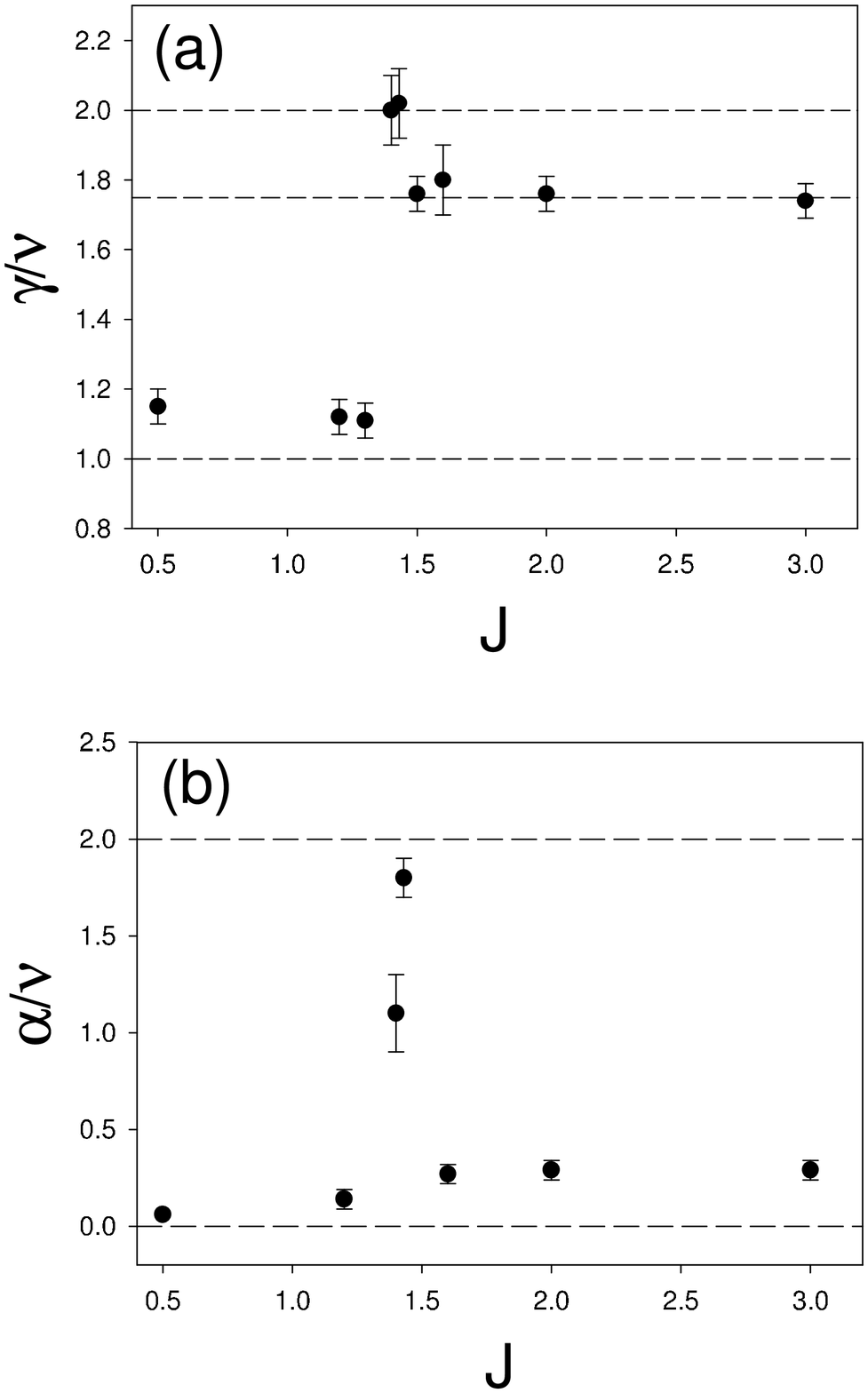}
\caption{\label{fig7} Critical exponents obtained from finite size
scaling as a function of $J$. (a) Susceptibility exponent
$\gamma/\nu$; (b) Specific heat exponent $\alpha/\nu$. The dashed
lines  indicate the reference values $1$, $1.75$ and $2$ in (a) and $0$ and $2$ in (b).}
\end{center}
\end{figure}

\begin{figure}
\begin{center}
\includegraphics[scale=0.37,angle=-90]{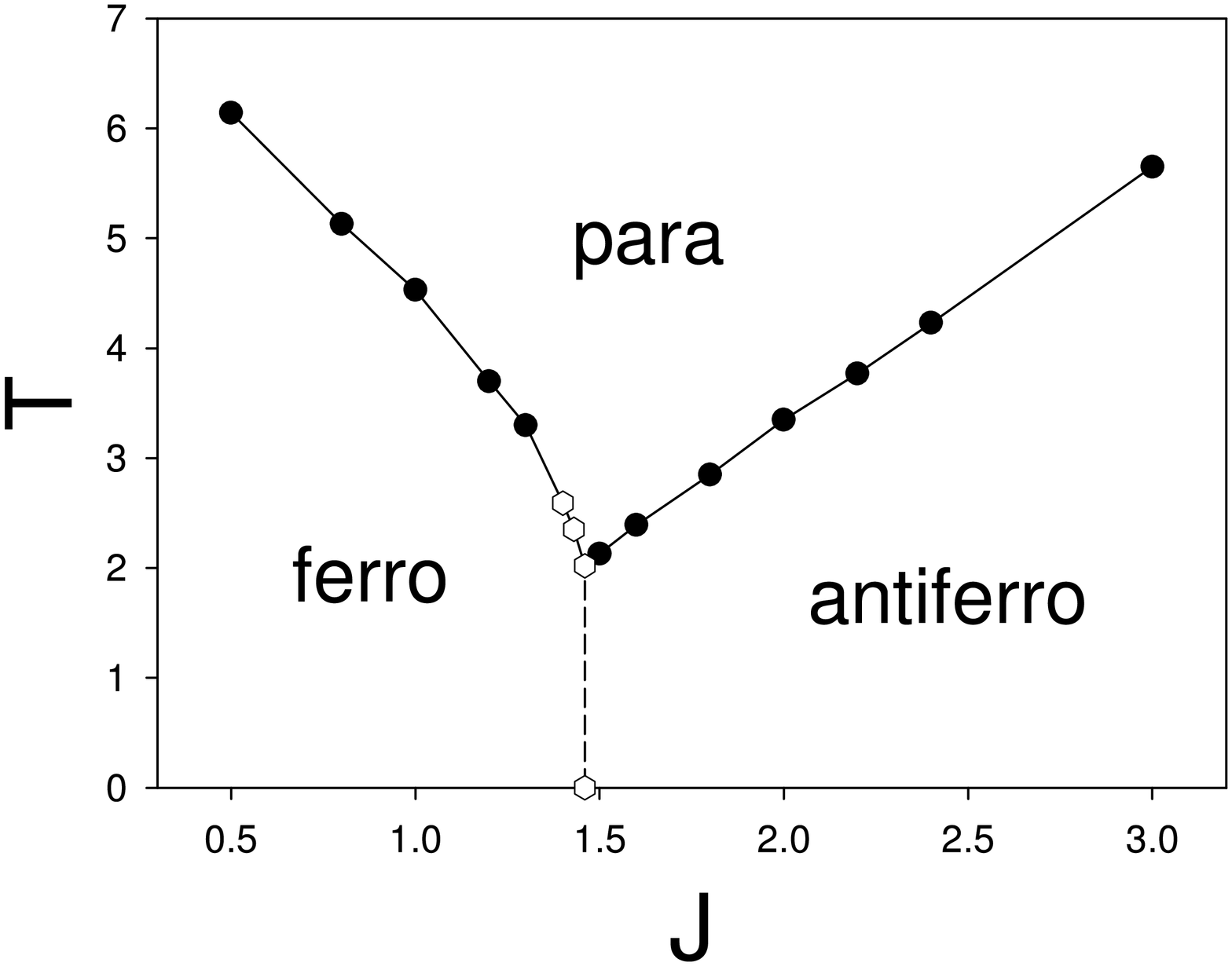}
\caption{\label{phased} Phase diagram $T$ vs. $J$. The critical
temperatures were estimated from the maxima of the specific heat.
Filled circles and open hexagons correspond to second and first
order phase transitions respectively.}
\end{center}
\end{figure}

\section{Non-equilibrium properties}

\label{nonequilibrium}

This section deals with the far-from equilibrium properties of the system
at low temperatures, i.e., its relaxation dynamics after a sudden
quench from $T=\infty$ to a temperature $T < T_c$.

\subsection{Non equilibrium domain structures: energy relaxation and characteristic domain length}

First  we  analyze the time evolution of the energy, with the time measured in MCS. We consider both the instantaneous energy per spin $E/N$ (i.e., the energy along single MC runs) and the mean excess of energy  $\delta e(t) \equiv \left[ H \right]/N-u(T)$, where $[\ldots]$ stands for average over different MC runs (i.e., over different realizations of the thermal noise).
 $u(T)$  is the equilibrium energy per spin at temperature $T$; $u(T)$ is obtained by equilibrating first the system during $10^4$ MCS starting from the ground state configuration and then averaging over a single MC run during $10^5$ MCS.

\begin{figure}
\begin{center}
\includegraphics[scale=0.35,angle=-90]{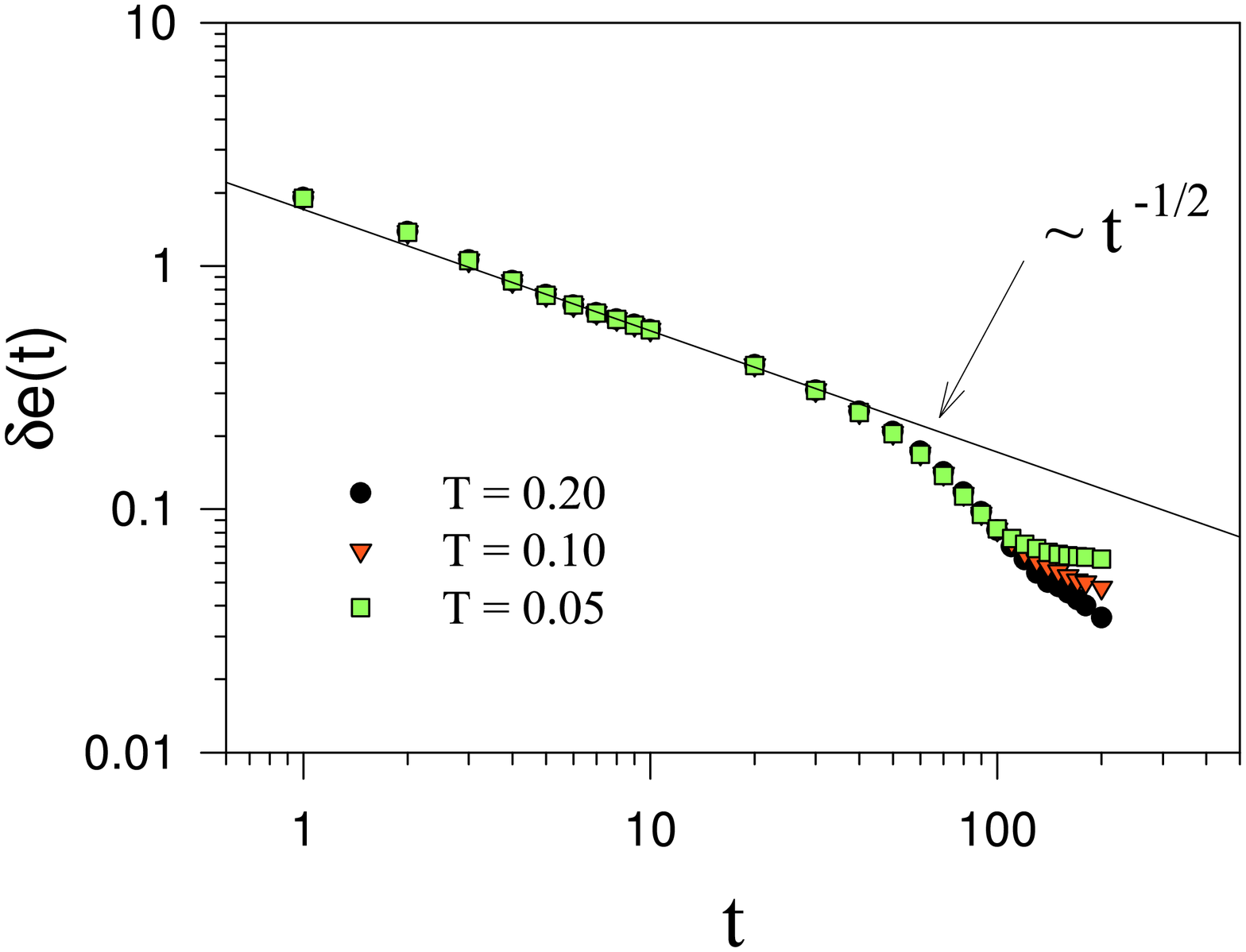}
\end{center}
\caption{ (Color on-line) Excess of energy $\delta e(t)$  as a
function of time for $J=1$, $L=100$  and different quench
temperatures $T<T_c$. The results were averaged over 2000 MC runs. \label{coar-ferro}}
\end{figure}

To check out our results we first calculate the evolution of $\delta e(t)$
in the simple case $J< J_t$ for different quench temperatures. The
typical behavior is shown in Figure \ref{coar-ferro}. We find that,
after a short transient period and before the system completely relaxes, the excess of energy behaves as  $\delta e(t)\sim t^{-1/2}$ independently of $T$. Since it is expected that $\delta e(t) \propto 1/l(t)$, where $l(t)$ is  the characteristic length scale of the domains, this behavior is consistent with a normal coarsening
process of a system with non-conserved order parameter\cite{Br2002}, where
$l(t)\sim t^{1/2}$.

\begin{figure}
\begin{center}
\includegraphics[scale=0.31,angle=-90]{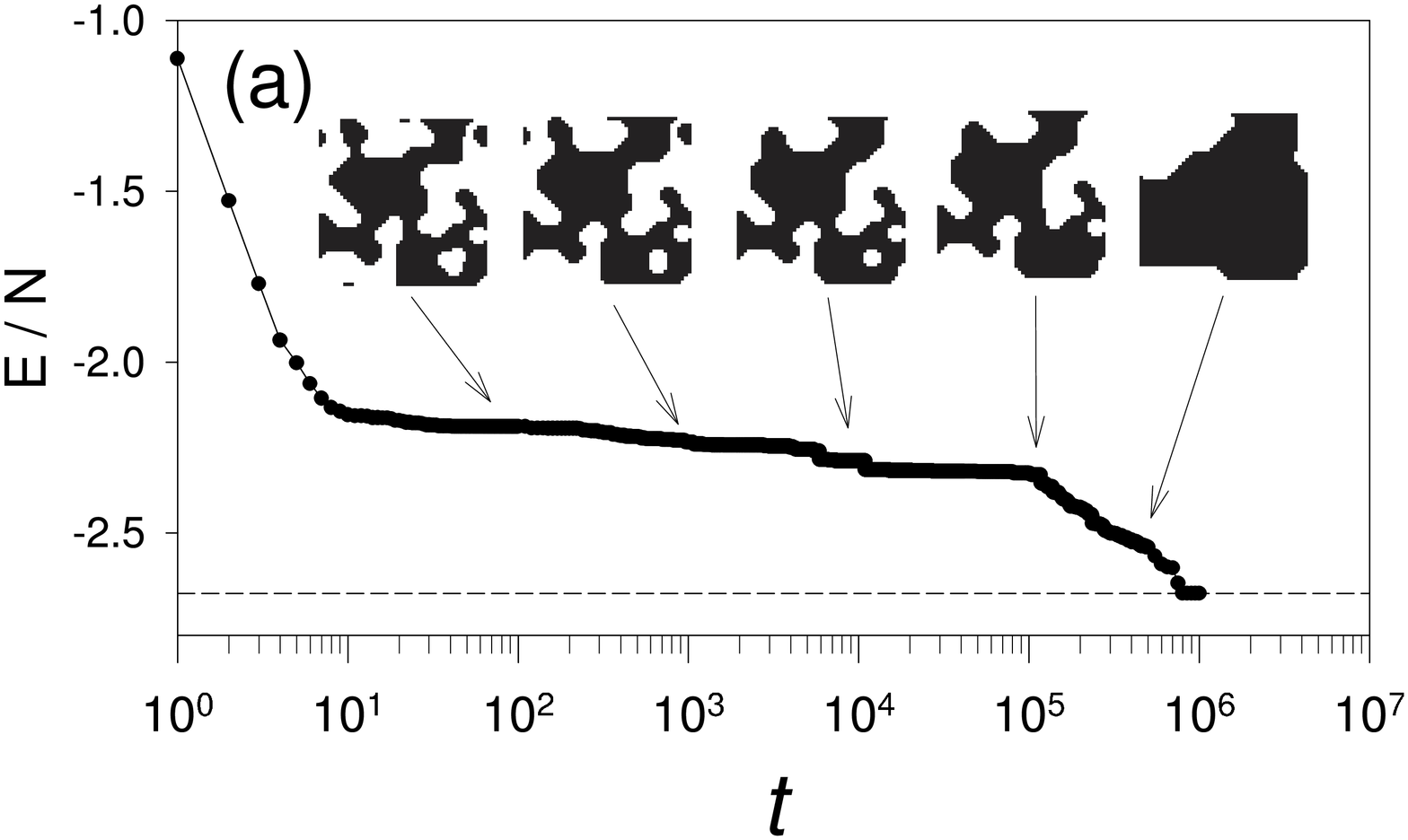}
\includegraphics[scale=0.31,angle=-90]{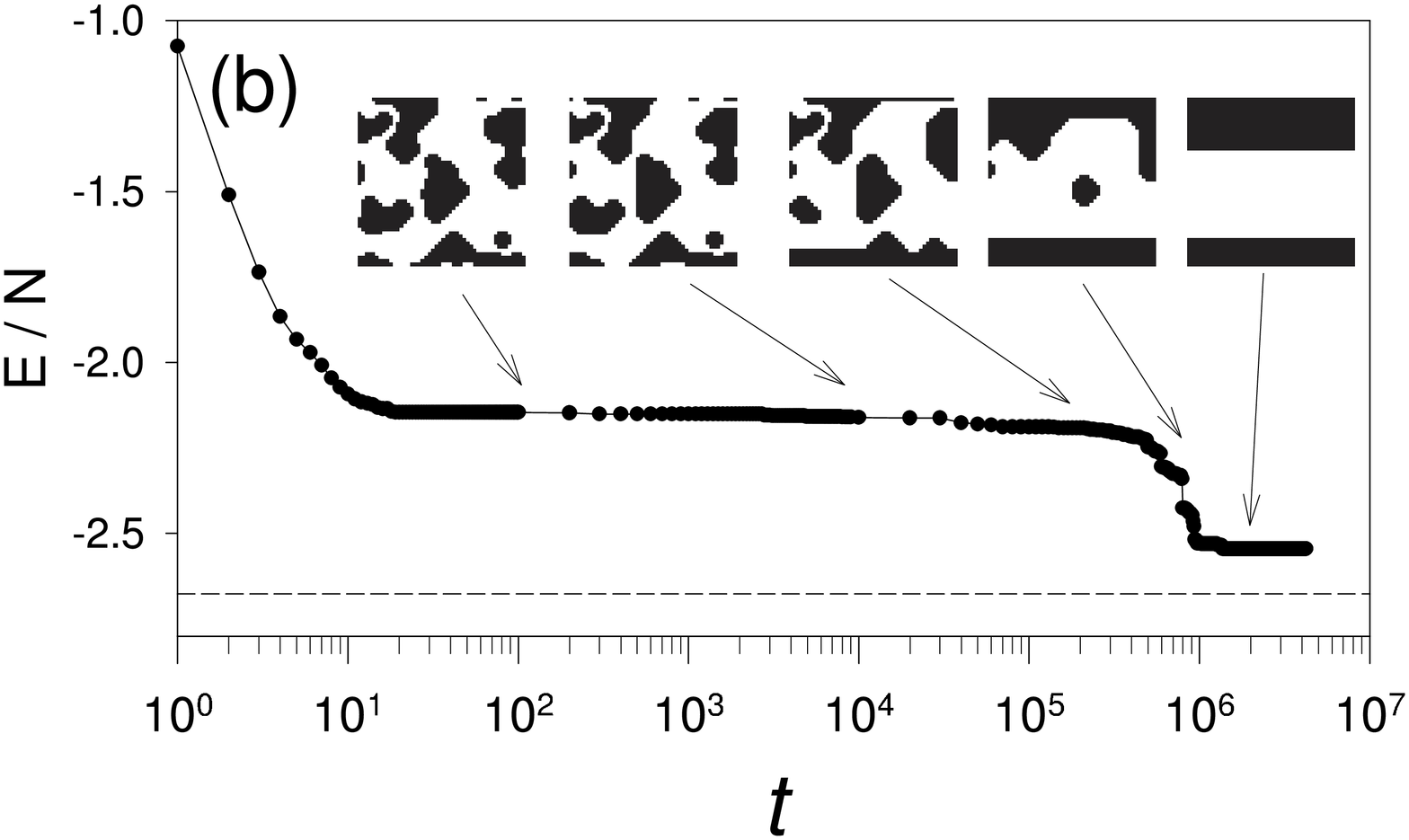}
\end{center}
\caption{Instantaneous energy per spin as a function of time for single realizations of the stochastic noise for $J=2$, $L=48$ and $T=0.04$. Typical antiferromagnetic domain configurations are shown along the evolutions, where black and white colors  codify regions with local staggered magnetization $m_s \approx 1$ and $m_s \approx -1$ respectively. The dashed lines correspond to the equilibrium energy at this temperature. (a) After living the glassy regime the system equilibrates. (b) After living the glassy regime the system gets stuck in a striped configuration.\label{energy}}
\end{figure}

\begin{figure}
\begin{center}
\includegraphics[scale=0.3,angle=-90]{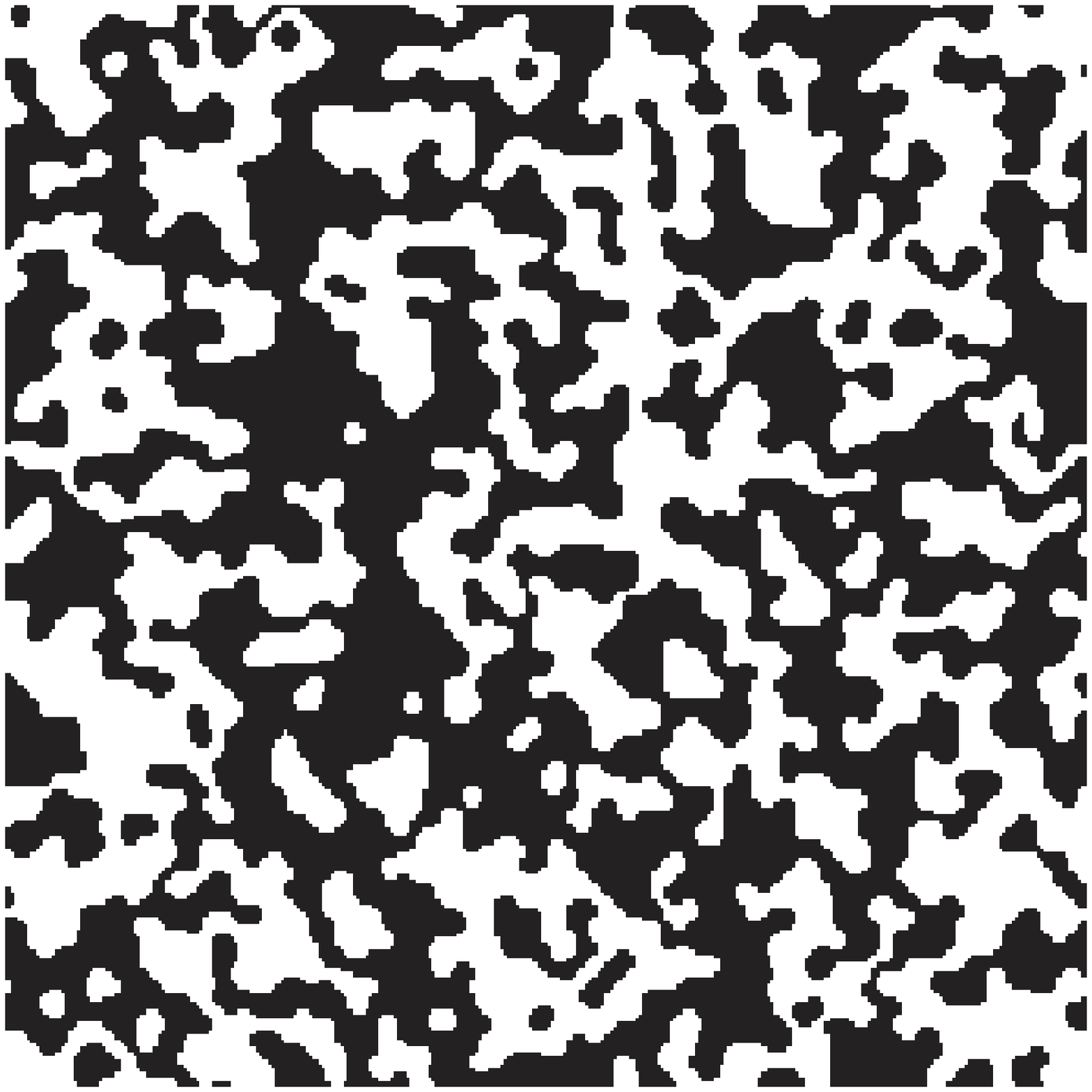}
\end{center}
\caption{Typical antiferromagnetic domain configuration for $J=2$, $T=0.04$, $L=256$ and $t=100;\; MCS$. Black and white follows the same convention as in Fig.\ref{energy}.  \label{energyconf}}
\end{figure}

Next, we  consider the relaxation in the antiferromagnetic
region  $J > J_t$ for different quench temperatures. At low enough temperatrues the relaxation of the system clearly departs from that expected in a normal coarsening process. The typical
behavior of the instantaneous energy is shown in Fig.\ref{energy}, together with   typical domain configurations along  single MC runs for $J=2$ and $T=0.04$. In that figure the domains correspond to regions of antiferromagnetic ordering, namely, black and white colors  codify regions with local staggered magnetization $m_s \approx 1$ and $m_s \approx -1$ respectively.

 Different relaxation regimes can be identified.
 After a short time quick relaxation process $0< t < \tau_0 \approx 20 \; MCS$, in which local antiferromagnetic order is set, the system always gets stuck in a complex non equilibrium disordered state composed mainly by a few intermingled  macroscopic antiferromagnetic domains; its typical shape is illustrated for a larger system size in Fig.\ref{energyconf}. This state presents a sort of labyrinth structure, in the sense that there is always at least one macroscopic connected domain, i.e., in such domain   any pair of points  can be connected by a continuous path without crossing a domain wall (see for example the black domain in Fig.\ref{energyconf}).  Up to certain characteristic time $\tau_1$ the system slowly relaxes  by eliminating small domains and fluctuations  located  in the large domain borders, in such a way that the local curvature  of the domain walls is reduced (Fig.\ref{energy}). Along this process the area of the main domains remains almost constant. In this sense, such process is reminiscent of a spinodal decomposition. We will call this the {\it glassy regime}. For time scales longer than a certain characteristic time $\tau_1$ both domains finally disentangle and relaxation is dominated by the competition between only two large domains, separated by rather smooth domain walls. Fig.\ref{energy} illustrates the two possible outcomes of this process: either the system relaxes directly to its equilibrium state (Fig.\ref{energy}a) or it gets stuck in an ordered configuration composed of   stripe shaped antiferromagnetic domains with almost flat domain walls (Fig.\ref{energy}b). We observe that both outcomes can happen with finite probabilities, the former being a bit more probable than the latter. The second case covers a large variety of configurations, including more than two stripes that can be oriented parallel to one of the coordinate axes (as in Fig.\ref{energy}b) or diagonally oriented (not shown). We will call this the {\it ordered regime}. Once the system arrives to one striped configuration, relaxation proceeds through the parallel movement of the domain walls, which perform a sort of random walk until two walls collapse and the system either attains the equilibrium state or gets stuck in a new striped configuration with a lesser  number of stripes. The mechanism of movement of the domain walls in this case is dent formation, i.e., single isolated spin flips along the interface creating an excitation that propagates along it, until either it disappears or covers the whole line\cite{SpKrRe2001}, which therefore advances in the perpendicular direction. The same kind of non-equilibrium structures and relaxation dynamics has been observed in  two dimensional short range interacting spin models at very low temperatures, namely the Ising\cite{SpKrRe2001} or Potts\cite{FeCa2007} models.  However, in those cases the movement of the dents are dominated by single spin flip barriers, while in the present one the associated mechanism is more complex due to the long range interactions.

\begin{figure}
\begin{center}
\includegraphics[scale=0.33,angle=-90]{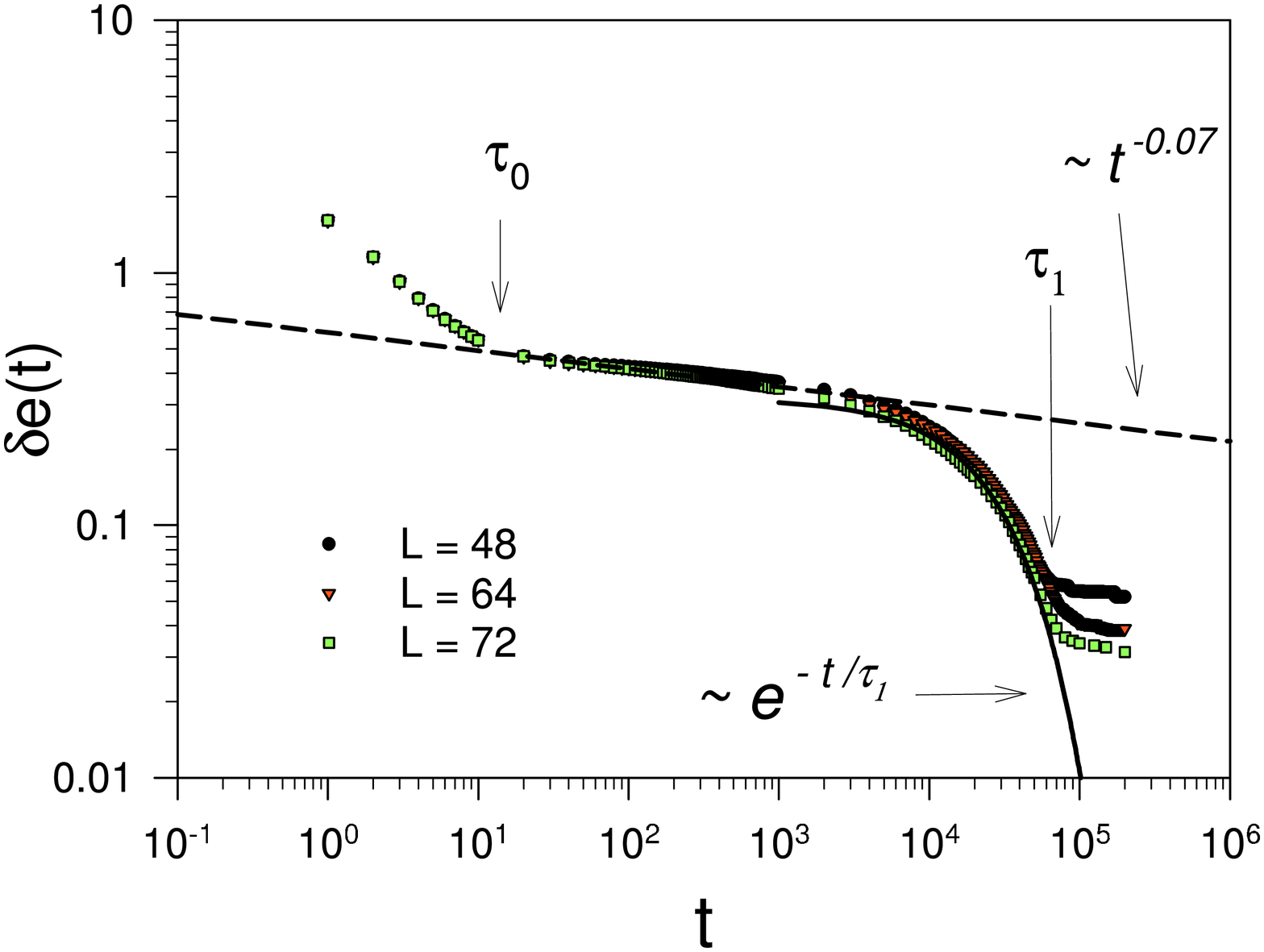}
\end{center} \caption{ (Color on-line) Excess of energy per spin (see text for details) as a function of time for $J=2$, $T=0.06$ and different system sizes.  Every curve was obtained by averaging over 400 runs. The dashed and full lines correspond to a power law and exponential fittings respectively.
\label{averenergiy1}}
\end{figure}

\begin{figure}
\begin{center}
\includegraphics[scale=0.36,angle=-90]{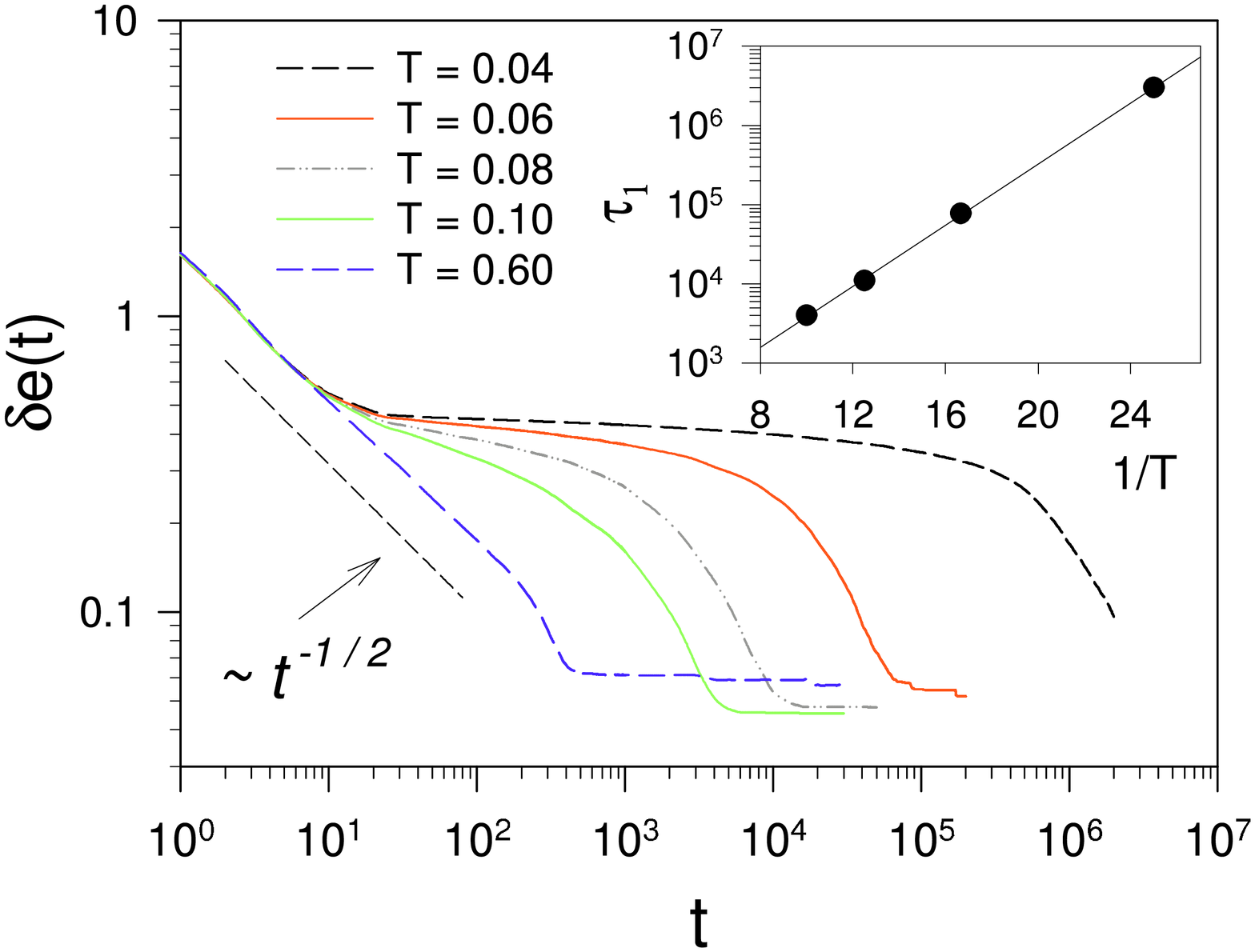}
\end{center} \caption{ (Color on-line) .
Excess of energy per spin  as a function of time for $J=2$, $L=48$ and different quench temperatures $T<T_c$ (decreasing from left to right). Every curve was obtained by averaging over 400 runs. The inset shows an Arrhenius plot of the crossover time. $\tau_1$.\label{averenergiy2}}
\end{figure}

In Fig.\ref{averenergiy1} we illustrate the typical behavior of the excess of energy $\delta e(t)$ at a fixed temperate for different system sizes. In Fig.\ref{averenergiy2} we show the excess of energy for different temperatures at a fixed system size. The three different relaxation regimes can be clearly seen in those curves: transient, glassy and ordered. The glassy regime appears for temperatures smaller than certain value $T_g$ ($T_g\approx 0.15$ for $J=2$). In this regime the excess of energy exhibits a size--independent pseudo--plateau, where it decays very slowly; indeed, the behavior of $\delta e(t)$ can be well  fitted by a power law $\delta e(t) \sim t^{-\omega}$,  with  very small exponents that decrease with temperature (the exponent for $J=2$ ranges from $\omega\approx 0.03$ for T=0.04 up to $\omega\approx 0.1$ for $T=0.1$), suggesting a logarithmic relaxation at very low temperatures. This suggests an activated dynamics with multiple energy barriers (we will return to this point later). After this regime, the system relaxes  exponentially into the ordered regime  $\delta e(t) \sim e^{-t/\tau_1(T)}$ (see Fig.\ref{averenergiy1}). The characteristic relaxation  time $\tau_1(T)$ can be estimated by fitting the corresponding part of the relaxation curve, as shown in Fig.\ref{averenergiy1}. The inset of  Fig.\ref{averenergiy2} shows an Arrhenius plot of $\tau_1$. The exponential decay, together with the clear Arrhenius behavior of $\tau_1$, indicates that the crossover between the two regimes is dominated by the activation through a single free energy barrier.

To gain further insight about the nature of the relaxation in the glassy regime, we analyze the scaling properties of the characteristic domain length $l(t)$. A sensible way to estimate the behavior of that quantity is to define it as\cite{ChGrGu1987,ShHoSe1992,LiJoEs2000}

\begin{equation}\label{ldet}
    l(t) \equiv \frac{-u(T)}{\delta e(t)}
\end{equation}

\begin{figure}
\begin{center}
\includegraphics[scale=0.5,]{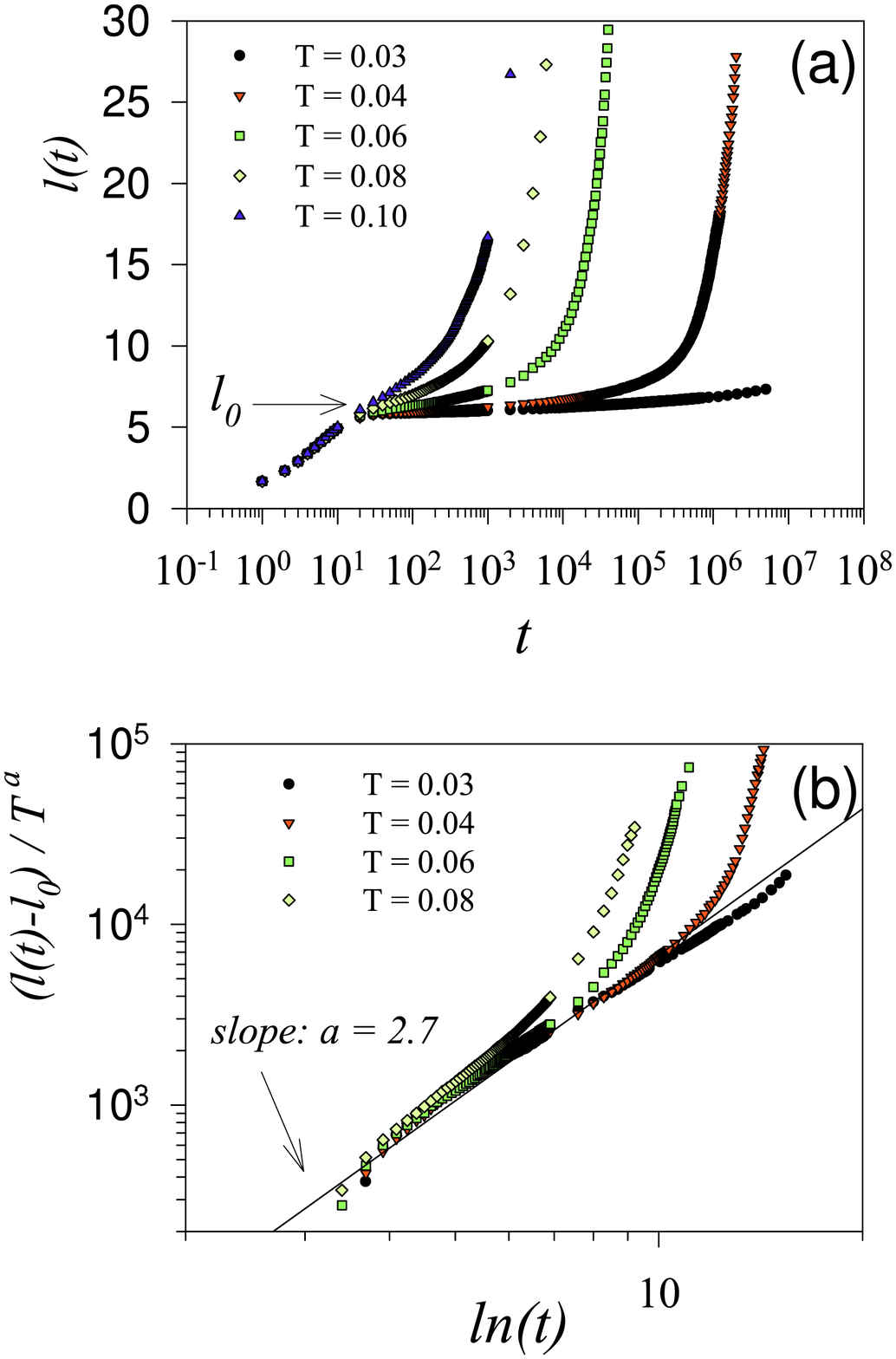}
\end{center} \caption{ (Color on-line) (a) Characteristic domain length  (see text for details) as a function of time for $J=2$, $L=48$ and different temperatures $T<T_c$ (decreasing from left to right). (b) Log-log plot of the normalized length $(l-l_0)/T^a$ vs. $ln(t)$ from the same data as in (a) for the lowest temperatures; $l_0 = 5.78$ is indicated in (a); the value of the exponent $a=2.7$ was chosen to obtain the best data collapse of the curves in the glassy regime. The straight line is a reference (power with exponent $a$).
\label{averlength}}
\end{figure}

In Fig.\ref{averlength} we show $l(t)$ for  $J=2$, $L=48$ and different temperatures $T<T_g$. The behavior of the excess of energy implies that, for time scales $\tau_0 < t< \tau_1$, $l(t)$ increases very slowly from a temperature-independent value $l_0=l(\tau_0)$; for time scales $t>\tau_1$ the characteristic length departs exponentially from the pseudo plateau (see Fig.\ref{averlength}a). We estimated $l_0$ as the average of the curves for different temperatures at $\tau_0$, obtaining $l_0\approx 5.78$. In Fig.\ref{averlength}b we show a double log plot of the rescaled quantity $(l(t)-l_0)/T^a$ vs. $ln(t)$. The exponent $a$ was chosen to obtain the best data collapse in the glassy regime of the data presented in Fig.\ref{averlength}a. Actually, a good data collapse inside the error bars of the statistical fluctuations is obtained for values of $a$ between 2.65 and 2.75; for values of the exponent outside that range the curves clearly do not collapse. Hence, we estimated $a=2.7 \pm 0.05$. The power law like behavior of the rescaled curves in Fig.\ref{averlength}b shows that the characteristic length behaves as

\begin{equation}\label{lenght}
    l(t) \sim l_0 + \left[ \frac{T}{b}\, \ln t \right]^a
\end{equation}

 \noindent for $t_0 < t < \tau_1(T)$ (a log-log plot of $l(t)-l_0$ vs. $t$ shows that a power law fit in the entire time interval is clearly inferior than in Fig.\ref{averlength}). Such behavior is consistent with a class 4 system, according to Lai et al classification\cite{LaMaVa1988}, i.e., a system with domain size dependent free energy barriers to coarsening\cite{ShHoSe1992} $f(l)$. In our case this would correspond to $f(l) \sim b\, (l-l_0)^{1/a}$. The numerical results suggest that in the present model such growth would stop when some maximum characteristic length $l_{max}$ is reached at $\tau_1(T)$, where the barrier becomes independent of $l$. After this point the system relaxes exponentially with a characteristic time $\tau_1 \propto \exp(F/T)$, where $F=f(l_{max})$ and therefore $l_{max} \approx l_0 +(F/b)^a$. From the data of Fig.\ref{averlength}b we estimate $b\approx 0.32$, while from the data of the inset of  Fig.\ref{averenergiy2} we estimated $F\approx 0.44$, giving an estimation $l_{max}\approx 8$.

\begin{figure}
\begin{center}
\includegraphics[scale=0.5]{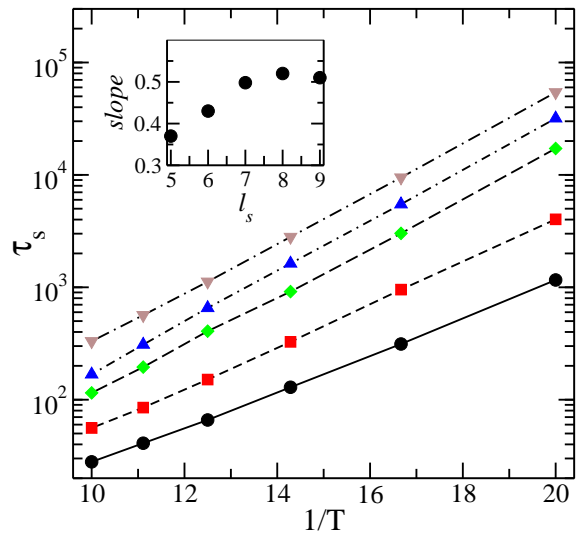}
\end{center} \caption{ (Color on-line) Arrhenius plot of the characteristic time for shrinking squares for different values of the square side $l_s$: from bottom to top $l_s=5,6,7,8,9$. The lines are a guide to the eye. The inset shows the energy barrier (slope of the linear fittings in the Arrhenius plot) as a function of $l_s$ (the symbol size in the inset is larger than the statistical error bars).
\label{shrinkingsquares}}
\end{figure}

To check the above interpretation we analyze the characteristic time $\tau_s$ for shrinking squares, i.e., the time needed for a square excitation of linear size $l_s$  to completely relax. This technique has been proved to be a sensitive way to check the relaxation dynamics of short range models when free energy barriers are involved\cite{ShHoSe1992,LiJo2000, LiJoEs2000}. In particular, Shore et al\cite{ShHoSe1992} have argued (and shown to be valid in particular cases) that the energy barriers to shrink square-shaped excitations should be a measure of the free energy barriers to coarsening. In our case we started with a ground state configuration of size $L$ with a square of inverted spins of size $l_s$ ($L \gg l_s$) and periodic boundary conditions. Although it is not clear to us whether the arguments of Shore et al\cite{ShHoSe1992} can be straightforwardly extended to  a system with long range interactions or not, one can still expect the barriers to shrink a square to provide at  least a rough measure of the free energy barriers to coarsening. In our case, this expectation is based on the direct observation of the domain configurations during relaxation in the glassy regime. We observe that rough domain walls tend to become flat rather fast, and that relaxation proceeds mainly at small jumps in the energy every time a sharp edge moves. The results for the time for shrinking squares support this conjecture. In Fig.\ref{shrinkingsquares} we show an Arrhenius plot of $\tau_s$ for different values of $l_s$ and temperatures $T<T_g$. We see that $\tau_s$ exhibits a clear Arrhenius behavior at all the temperatures for $l_s>4$ (for  sizes $l_s \leq 4$ the squares shrink quickly in a few MCS), with associated barriers that grow slowly for $l_s < 8$ and saturate for $l_s\geq 9$ at a value around $0.5$, close to  $F=0.44$. Although the limited range of values of $l_s$ where the barrier shows a dependency on it does not allow a more accurate comparison, the consistency with the previous interpretation of the behavior of $l(t)$ is clear.

For temperatures larger  than $T_g$  the glassy regime completely disappears and the system decays through a normal coarsening process, i.e.,  $\delta e(t)\sim t^{-1/2}$ (see  Fig.\ref{averenergiy2}). However, for some range of temperatures it still gets stuck in some  long-lasting antiferromagnetic striped  configuration with high probability, so the corresponding plateau in the excess of energy is still observable (for $J=2$ we observed it for temperatures up to $T\approx 1.5$). Those configurations are highly stable, even at relatively high temperatures. The characteristic equilibration time $\tau_2$, defined as the time after which the system attains the equilibrium state with probability one, is very difficult to estimate, but it is at least three orders of magnitude larger than $\tau_1$ for $T<T_g$.

\subsection{Time correlation and response functions}

Another way to characterize the out of equilibrium
dynamics of complex magnetic systems is through the analysis of
the two-time autocorrelation function $C(t,t^\prime)$. A system
that has attained thermodynamical equilibrium or meta--equilibrium satisfies time
translational invariance (TTI), i.e.,  $C(t,t^\prime) \equiv
C(t-t^\prime)$, al least for certain time scales. Far from equilibrium TTI is broken and time
correlations exhibits a dependency on the history of the sample
after the quench. This phenomenon is called {\it aging} and in
real systems it can be observed through a variety of experiments.
A typical example is the zero-field-cooling\cite{LuSvNoBe1983}
experiment, in which the sample is cooled in zero field to a
subcritical temperature at time $t=0$. After a waiting time $t_w$
a small constant magnetic field is applied and the time evolution
of the magnetization is recorded. It is then observed that the
longer the waiting time $t_w$ the slower the relaxation and this
is the origin of the term {\it aging}. Moreover, the scaling
properties of two-times quantities provides information about
the underlying relaxation dynamics\cite{ViHaOcBoCu1997,HePlS2007}.

\begin{figure}
\begin{center}
\includegraphics[scale=0.52]{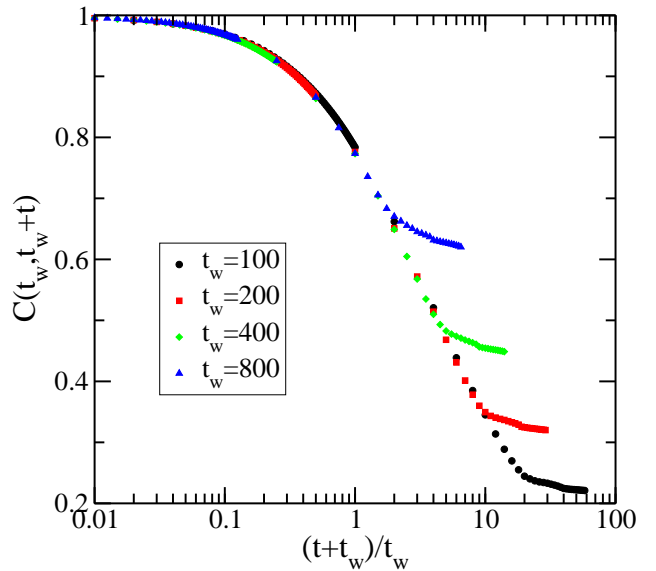}
\end{center} \caption{ (Color on-line) Two-times autocorrelation function $C(t_w+t,t_w)$ as a function of $(t+t_w)/t_w$ for $J=1$ (ferromagnetic phase), T=0.2, $L=600$ and different values of the waiting time $t_w$ (increasing from bottom to top).
\label{coarsening-ferro}}
\end{figure}

Although aging can be detected through different time-dependent
quantities, a straightforward way to establish it in a numerical
simulation is to calculate the spin autocorrelation function

\begin{equation}
C(t_w+t,t_w) = \left[ \frac{1}{N}\, \sum_{i=1}^{N} \sigma_i(t_w+t)
\sigma_i(t_w) \right],
\end{equation}
where  $t_w$ is the waiting time from the quench at $t=0$
(completely disordered initial state) and $\left\lbrace
\sigma_i(t)\right\rbrace $ is the spin configuration at time $t$.

First of all we calculate $C(t_w+t,t_w)$ in the ferromagnetic
part of the phase diagram, i.e., for $J<J_t$. We found that
$C(t_w+t,t_w)$ depends on $t$ and $t_w$ through the ratio $t/t_w$, as shown in Fig.\ref{coarsening-ferro}.
This type of scaling is called {\it simple aging} and it is
characteristic of a simple coarsening (i.e., domain growth)
process. This result is in agreement with the observed behavior of the excess of energy (Fig.\ref{coar-ferro})

\begin{figure}
\begin{center}
\includegraphics[scale=0.55,angle=0]{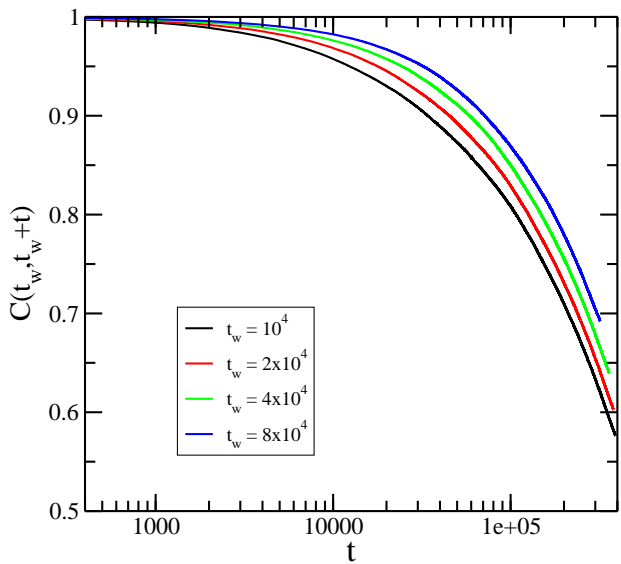}
\end{center}
\caption{(Color online) Two-times autocorrelation function as a function of the
difference of times $t$ for $J=2$, $T=0.04$, $L=256$ and different
values of the waiting time $t_w$ (increasing from left to right).\label{agingfig1}}
\end{figure}

Next we consider the behavior of the correlations during the glassy regime observed in the previous section for $J>J_t$.
 The typical behavior of the autocorrelation function is
shown in Figure~\ref{agingfig1}, where we plot $C(t_w+t,t_w)$ {\em
vs} $t$  for $T=0.04$, $J=2$ and $L=256$ and different waiting times. The
simulation was run up to $t=5\times 10^5$ MCS and typical averages
were performed over $2000$ realizations of the thermal noise; both times were chosen such that $\tau_0<t+t_w<\tau_1$.  A
first trial to collapse those curves showed that in this case
$C(t_w+t,t_w)$ does not exhibit simple aging. A similar behavior is observed for $J=2$, $T=0.06 <T_g$ and $\tau_0<t+t_w<\tau_1$.

\begin{figure}
\includegraphics[scale=0.35,angle=-90]{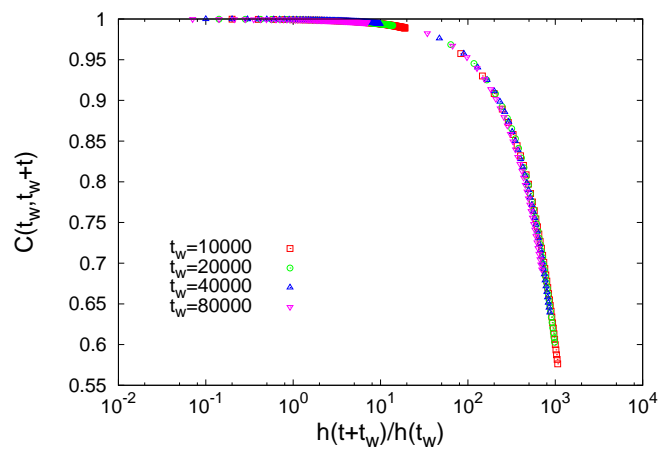}
\caption{\label{agingfig2} (Color on-line) Data collapse of the
autocorrelation curves of Fig.\ref{agingfig1} ($J=2$, $T=0.04$ and $L=256$) using the scaling function Eq.(\ref{scalfunc}) with scaling parameters shown in Table \ref{table1}.}
\end{figure}

It has been  observed for a large variety of systems that in the
aging scenario the curves of $C(t_w+t,t_w)$ for different $t_w$
always collapse into a single one using an adequate {\em scaling
function}\cite{ViHaOcBoCu1997}. Although there is no theoretical basis for
determining the scaling function, there are a few choices that
have been able to take into account both experimental and
numerical data, perhaps the most frequent one being the additive form
\begin{equation}\label{scalcorre}
C(t_w+t,t_w) = C_{st}(t) + C_{ag}\left(
\frac{h(t_w+t)}{h(t_w)}\right)  .
\end{equation}

\noindent where $C_{st}(t)$ is a stationary part, usually well described by an algebraic decay
\begin{equation}
C_{st}(t) = B\, t^{-\gamma}.
\end{equation}
\noindent The function $h$, appearing in the aging
part of the autocorrelations $C_{ag}$, is some scaling function
(in the case of simple aging $h(t)$ is a power law that describes
the characteristic linear domain size growth). In our case the
best data collapse of the autocorrelation curves was obtained
 using a scaling function of the form
\begin{equation}\label{scalfunc}
h(t) = \exp\left[ \frac{1}{1-\mu}\left( \frac{t}{\tau}
\right)^{1-\mu}\right],
\end{equation}
\noindent which has been used to account for experimental data\cite{ViHaOcBoCu1997} and
in the Edwards--Anderson model for spin glasses \cite{StMoTa2003},
where  $\tau$ is a microscopic time scale. It is worth to note
that the scaling function (\ref{scalfunc}) interpolates a range of
scenarios: from {\em sub--aging} for $0<\mu<1$, to {\em
super--aging} for $\mu>1$, through {\em simple aging} for\cite{ViHaOcBoCu1997} $\mu=1$; for $\mu=0$ TTI is recovered.
 In Figure~\ref{agingfig2} we
 show  the collapse of the data from
Figure~\ref{agingfig1} using the scaling parameter values shown in Table \ref{table1} ($\tau$ was arbitrarily fixed  to one). A similar data collapse was observed for $J=2$ and $T=0.06$ (see parameters in Table \ref{table1}). To check possible finite size effects we performed a similar calculation for $T=0.04$, $J=2$ and $L=64$, finding the same collapse shown for $L=256$ in Figure~\ref{agingfig2} with the same scaling parameters; only a small variation in the master curve is observed. We see that for $T<T_g$ the best data collapse is obtained without stationary part and a clear sub-aging is observed.

\begin{table}
\begin{center}
\begin{tabular}{|c|c|c|c|}
\hline
$T$ & $B$ & $\gamma$ & $\mu$ \\
\hline
 $0.04$ & $0$ & - & $0.50$\\
\hline
 $0.06$ & $0$ & - & $0.30$\\
\hline
 $0.6$ & $0.19$ & $0.22$ & $0.25$\\
\hline
 $1.0$ & $0.16$ & $0.30$ & $0.14$\\
\hline
\end{tabular}
\end{center}
\caption{Scaling parameter values obtained from the best data collapse for the correlation curves for $J=2$ at different temperatures  using the scaling forms Eqs.(\ref{scalcorre})-(\ref{scalfunc}).}
\label{table1}
\end{table}

We also repeat the correlation calculation for temperatures $T>T_g$ at time scales corresponding to the ordered regime $t_w+t > \tau_1$($T=0.6$ and $T=1$ for $J=2$ and $L=64$). Again, aging is observed in this regime and a  data collapse similar to that shown in Figure~\ref{agingfig2} using the scalings (\ref{scalcorre})-(\ref{scalfunc}) is obtained. The corresponding scaling parameter values are shown in Table \ref{table1}. We see that for this temperature range the best data collapse is obtained by including a stationary part and that the scaling parameter $\mu$ decreases systematically as the quench temperature increases  signaling that the system approaches  TTI. We find that $\mu$ becomes zero at a temperature $T \approx 1.5 < T_c$, which can be considered as the onset of this non exponential relaxation.

To further characterize this non equilibrium behavior we also analyze the generalized Fluctuation-Dissipation
Relations (FDR), which can be expressed as  \cite{CuKu1993}:

\begin{equation}
R(t_w+t,t_w) = \frac{X(t_w+t,t_w)}{T} \frac{\partial C(t_w+t,t_w)}
{\partial t_w}
\end{equation}

\noindent where $R(t_w+t,t_w)= 1/N \; \sum_i \partial \left< \sigma_i(t_w+t) \right>/\partial h_i(t_w)$ is the response  to a local
external magnetic field $h_i(t)$  and $X(t_w+t,t_w)$ is the
fluctuation dissipation factor. In equilibrium the Fluctuation Dissipation Theorem (FDT) holds and $X(t_w+t,t_w)= 1$, while out of
equilibrium $X$ depends  on $t$ and $t_w$ in a non trivial way. It has been conjectured \cite{CuKu1993} that
 $X(t_w+t,t_w)= X[C(t_w+t,t_w)]$. This conjecture has proved valid in all systems studied to date.

Instead of considering the response function it is easier to analyze
the integrated response function
\begin{equation}
\chi(t_w+t,t_w)=\int_{t_w}^{t_w+t} R(t_w+t,s)\,  ds.
\end{equation}
Assuming $X(t_w+t,t_w)= X[C(t_w+t,t_w)]$ one obtains
\begin{equation}
 T\chi(t_w+t,t_w)= \int_{C(t_w+t,t_w)}^1 X(C) dC
\end{equation}
and by plotting $ T\, \chi$ vs. $C$ one can extract $X$ from the slope of the curve.
In particular, if the FDT holds $X=1$ and $T\chi(t) = (1-C(t))$; any departure from
this straight line  brings information about the non-equilibrium process.  In numerical simulations of spin glass
\cite{PaRiRu1998}, structural glass\cite{Pa1997} and Random Anisotropy Heisenberg\cite{BiCaTa2005}  models it has
been found that, in the non-equilibrium regime, this curve follows another straight
line with smaller (in absolute value) slope when $t/t_w \gg 1$. In this case the FD factor $X$ can
be interpreted in terms of an effective temperature \cite{CuKuPe1997} $T_{eff}=T/X$.

We  apply this procedure  during the glassy and ordered relaxation regimes previously found for $J > J_t$. At time $t_w$ we took a copy of the system spin configuration, to which a random
magnetic field $h_i(t)=h\, \epsilon_i$ was applied, in order to avoid favoring
long range order\cite{Ba1998,StCa1999}; $\epsilon_i$ was taken from a bimodal distribution
($\epsilon_i=\pm 1$). We have
used different values of $h$ in order to check that the system was within
the linear response regime. All the results presented here were obtained with $h=0.025$.

\begin{figure}
\begin{center}
\includegraphics[scale=0.5,angle=0]{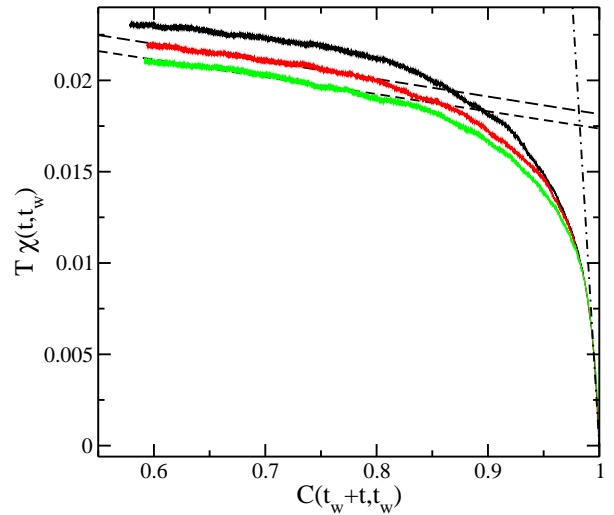}
\end{center}
\caption{Parametric plot of $T \chi(t,t_w)$ vs $C(t_w+t,t_w)$ in the glassy regime
 ($J~=~2$, $T=0.04$ and $L=256$;$t_w+t<\tau_1$) for different waiting times. The dashed lines are linear fittings; the dash-dotted lines represents the  thermal equilibrium relation $T \chi = 1-C$. The linear fittings have a slope $X=0.01$, corresponding to an effective temperature $T_{eff}=4$. Waiting times from top to bottom are $8\times 10^4$, $10^5$ and $2\times 10^5$ MCS respectively.\label{fduno}}
\end{figure}

In Fig.~\ref{fduno} we display  $T \chi(t,t_w)$ vs $C(t_w+t,t_w)$ in a parametric plot for different waiting times in the glassy regime (i.e., $t_w+t<\tau_1$) at  $T=0.04$. We observe a typical two time separation
behavior\cite{Cu2002}. At $t=0$ the system starts
in the right bottom corner (fully correlated and demagnetized)
and during certain time (that depends on $t_w$) it follows the equilibrium straight line, indicating
the existence of a quasi equilibrium regime. Nevertheless, at
certain time the system clearly departs from this quasi--equilibrium
curve and moves along a different straight line, but with a different
(smaller) slope, indicating an effective temperature that is larger
than the temperature of the thermal bath ($T_{eff}=4$ for the data of Fig.\ref{fduno}).  Notice that  the quasi equilibrium regime is very small, consistently with the absence of a stationary part observed in the correlation function (see Table \ref{table1}).

\begin{figure}
\begin{center}
\includegraphics[scale=0.5,angle=0]{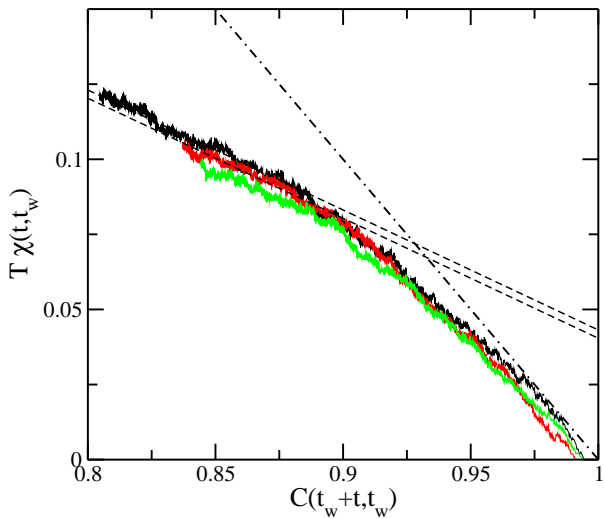}
\end{center}
\caption{Parametric plot of $T \chi(t,t_w)$ vs $C(t_w+t,t_w)$ in the ordered regime
 ($J~=~2$,  $T=0.6$ and $L=64$; $t_w+t>\tau_1$)  for different waiting times. The dashed lines are linear fittings; the dash-dotted lines represents the  thermal equilibrium relation $T \chi = 1-C$. The linear fittings have a slope $X=0.4$, corresponding to an effective temperature $T_{eff}=1.5$. Waiting times from top to bottom are $4\times 10^4$ , $8\times 10^4$ and $10^5$ MCS respectively.\label{fddos}}
\end{figure}

In Fig.~\ref{fddos} we display  $T \chi(t,t_w)$ vs $C(t_w+t,t_w)$ in a parametric plot for different waiting times in the ordered regime (i.e., $t_w+t>\tau_1$) at  $T=0.6$. The two--slope structure is again observed, although with a smaller effective temperature ($T_{eff}=1.5$).

\section{Discussion}

We introduced a lattice spin model that mimics a system of interacting particles through a short range repulsive potential and a long range power law decaying potential. Through a detailed Monte Carlo simulation analysis we computed the complete equilibrium phase diagram of the model at finite temperature and characterized the order of the different transition lines. We showed that the model presents only two simple ordered phases at low temperatures: ferromagnetic (for $J<J_t$) and antiferromagnetic (for $J>J_t$), without any trace of geometrical frustration and/or complex patterns.

We then analyzed the out of equilibrium relaxation of the system after a quench from infinite temperature down to subcritical temperatures, in different regions of the phase diagram. While a normal coarsening  behavior appeared in the ferromagnetic region of the phase diagram $J<J_t$ (i.e., a domain growth process that follows Allen-Cahn law $l(t)\sim t^{1/2}$) the system shows a complex relaxation scenario in the antiferromagnetic region $J>J_t$. This is precisely the most interesting situation, since for those values of $J$ the pair interaction potential of the model shows the same qualitative features as a continuous, Lennard-Jones (LJ) like potential, namely, a nearest neighbors repulsive interaction (i.e, ''hard core'' like) and an attractive, power law decaying interaction at longer distances (see Fig.\ref{fig1}). We must stress that we did  not intend to present a lattice version of the LJ gas, but to show that some very basic features present in it (i.e., the competition between short and long range interactions)  are enough to produce non trivial slow relaxation properties.  We observed that
such competition gives rise to  non--equilibrium structures that strongly slow down the dynamics, even in the absence of geometrical frustration and/or imposed disorder. The most interesting of those structures  give rise to a relaxation regime with several appealing properties that strongly resemble those observed in different glassy systems.

First of all, that regime is characterized by a dynamically generated disordered non-equilibrium state, characterized by a labyrinth structure, i.e., composed of at least  one macroscopic connected domain. The energy of such state displays a pseudo-plateau and a finite lifetime $\tau_1$ that diverges when $T \to 0$; both properties are independent of the system size. Such phenomenology is extremely reminiscent of transient particles colloidal gels obtained by quenching a monodisperse gas of colloidal particles under Brownian Dynamics (Langevin dynamics in the overdamped limit) that interact through  a generalized $2n-n$ LJ potentials\cite{LoHe1999,Di2000}. A gel is a non--equilibrium disordered state, characterized  as a percolating cluster of dense regions of particles with voids that coarsen up to certain size and freeze when the gel is formed; transient gels do not have permanent bonds between them and collapse  after a finite life time\cite{Di2000}. The energy of transient gels of $2n-n$ LJ colloidal particles displays a slowly decaying pseudo plateau\cite{Ca2009}, as in the present case. It has also been observed that the life time of the gel strongly increases as the interaction range is decreased\cite{Ca2009}, for instance, by changing the value of $n$. It would be interesting to check if a similar effect can be  obtained in the present model by changing the interaction range (for instance, by changing the exponent of the long range interaction term in Hamiltonian (\ref{Hamilton2})).

Second,  during the glassy regime the dynamics appears to be governed by free energy barriers to coarsening that scale as a power law  with the characteristic domain size, with an associated  logarithmic growth $l(t) \sim [T\, \ln(t)]^a$ (and therefore a logarithmic relaxation).  Such behavior is characteristic of  class 4 systems, according to Lai et al classification\cite{LaMaVa1988}. Some  examples of non--disordered short--range interacting systems that present growing free energy barriers with the domain size are already known, such as the three dimensional Shore and Sethna (SS) model\cite{ShSe1991,ShHoSe1992,Kr2005} (i.e., an Ising model with nearest neighbors ferromagnetic interactions and next nearest neighbors antiferromagnetic interactions) and a generalization of the previous one introduced by Lipowski et al\cite{LiJoEs2000}, including a four spin plaquette interaction term. However, in those models the barriers appears to grow linearly with the domain size (which corresponds to a pure logarithmic growth $l(t) \sim \ln(t)$) and therefore fall into the class 3 category of Lai et al\cite{ShHoSe1992}. So far, examples of class 4 systems were found only among disordered systems, such as spin glasses\cite{FiHu1986} and the Ising model with random quenched impurities\cite{HuHe1985}. To the best of our knowledge, this would be the first possible example of class 4 behavior in a non disordered  system. Nevertheless,  an important difference between the above mentioned non--disordered models and the present one have to be remarked. While in those models the logarithmic growth appears to lead to divergent barriers, in the present case the apparently barriers  growth stops at some maximum value $F$ that determines the life time $\tau_1$. This implies the existence of a characteristic length $l_{max}$ in the dynamics of the system. Although such limited length scales makes very difficult to obtain better numerical evidence of the existence of growing barriers, the consistency between the scaling of the excess of energy and the time for shrinking square excitations gives support to our conjecture. Thus this system appears to behave, at least for certain time scale that can become very long a very low temperatures, as a class 4 system, even though in the long term it behaves as a class 2 system in the sense that its ultimate dynamics is governed by a single free energy barrier. This opens the possibility of having a truly non disordered class 4 system if the lifetime of the glassy state at finite temperature could be extended by tuning the range of the interactions, as previously discussed.
Indeed, we believe that the possibility of having in a non--disordered class 4 system  makes it worth to further investigation.

While we do not have an explanation for such possible relaxation scenario (i.e., power law growth of barriers with a rather small exponent $1/a$ and limited length scale for growing) probably a key ingredient to explain it would be the moderated long range character of the interactions. It would be very interesting to check if the same behavior can be detected in the LJ gas or its generalizations, which appear to present a similar phenomenology\cite{Ca2009}. However, the range interactions would be not enough in the present model to generate dynamical frustration (in the sense stated by Shore et al\cite{ShHoSe1992}, i.e., systems whose dynamics is slowed down by the presence of  growing free energy barriers), but the type of competition between interaction would be equally important. This can be clearly  seen by looking at the non--equilibrium behavior after a quench of the reverse model, namely, that given by the Hamiltonian (\ref{Hamilton1}) with the inverse coefficients sign ($J_1<0$ and $J_2>0$). While the equilibrium phase diagram of that model is by far more complex than  the present one\cite{CaMiStTa2006,PiCa2007},  its domain growth behavior after a quench to subcritical temperatures is relatively simple, at least for the regions of the phase diagram explored up to now. Depending on the ratio of couplings $J_1/J_2$ it behaves as a class 2 system\cite{GlTacaMo2003} (which implies domain--size independent free energy barriers), like the 2D SS model\cite{ShSe1991,ShHoSe1992} , or relaxation can be dominated by nucleation effects\cite{CaMiStTa2008}. Accordingly, that system presents simple aging \cite{ToTaCa1998,GlMo2006} and trivial FD relations\cite{StCa1999} (infinite effective temperature), at variance with the present case.

In the glassy regime of the present model we found non trivial aging effects, with scaling properties characteristic of glassy systems (subaging). Non trivial aging has also been found in the non disordered four-spin ferromagnetic model\cite{SwBoTrBr2000} (a particular case of the model of Lipowski et al\cite{LiJoEs2000}), but in this case the system displays superaging, while the disordered version of the same model displays subaging\cite{AlFrRi1996}. Subaging has also been reported in molecular dynamic simulations of small LJ clusters\cite{OsSeTa2002}.

We also found FD relations displaying a well defined effective temperature in the aging regime. It is interesting to note that, although non trivial behavior of time correlations and responses are usually associated to glassy systems, they have also been observed in  very simple systems which undergo domain growth at intermediate time scales (as in the present case), namely, the ferromagnetic Ising chain\cite{LiZa2000} and the 2D ferromagnetic Ising and Potts models with Kawasaki dynamics\cite{Kr2005}. A similar behavior has been found in a non-disordered plaquette model for glasses introduced by Cavagna et al\cite{CaGiGr2003a,CaGiGr2003b}. It is worth noting that the 3D SS model presents only trivial FD relations, even for the temperature range where it appears to present logarithmic growth of domains\cite{Kr2005}.

 \label{conclusions}

 This work was partially supported by grants from CONICET,
 SeCyT-Universidad Nacional de C\'ordoba
and FONCyT grant PICT-2005 33305 (Argentina). We thank D. Stariolo for critical reading of the manuscript.

\bibliographystyle{apsrev}


\end{document}